\documentstyle[12pt,aaspp4]{article}
\newcommand\etal{{\it et al}. }
\newcommand\be {\begin{equation}}
\newcommand\en{\end{equation}}

\slugcomment{Submitted to {\it The Astrophysical Journal}}

\received{2003 October 14}
\begin{document}

\title{MODELS FOR THE GRAVITATIONAL FIELD OF THE GALACTIC BAR. AN 
APPLICATION TO STELLAR ORBITS IN THE GALACTIC PLANE AND ORBITS OF 
SOME GLOBULAR CLUSTERS}

\author{B\'arbara Pichardo \altaffilmark{1,}\altaffilmark{2}, 
Marco Martos \altaffilmark{1}, Edmundo Moreno \altaffilmark{1}}

\altaffiltext{1}{Instituto de Astronom\'\i a, Universidad Nacional 
Aut\'onoma de M\'exico, A.P. 70-264, 04510 M\'exico D.F., M\'exico;
Electronic mail: barbara@astroscu.unam.mx, marco@astroscu.unam.mx;
edmundo@astroscu.unam.mx}
\altaffiltext{2}{Astronomy Department, University of Wisconsin at 
Madison, 475 N.Charter Street, Madison WI 53706, USA.
Electronic mail: barbara@astro.wisc.edu}

%%%%%%%%%%%%%%%%%%%%%%%%%%%%%%%%%%%%%%%%%%%%%%%%%%%%%%%%%%%

\begin{abstract}

We built three models for the gravitational field of the Galactic bar.
These models are an inhomogeneous ellipsoid, an inhomogeneous prolate
spheroid, and a superposition of four inhomogeneous ellipsoids. Among
the three models,  the superposition provides our best approximation
to the observed boxy mass distribution of the Galactic bar. Adding the
bar component to an axisymmetric Galactic model, we have calculated
stellar midplane orbits and orbits of some globular clusters with
known kinematical data. For all models we find a secular dispersion
effect upon the orbital energy and angular momentum, as measured in
the Galactic inertial  frame. This effect might be relevant to explain
the orbital  prograde-retrograde distribution of globular
clusters. For the stellar  kinematics, we study the connection between
the sense of orbital motion  in the midplane and the onset of chaos in
the presence of the bar. In the inner region of the bar, chaos is
induced by an axisymmetric central component (bulge) and it arises in
orbits that change its orbital sense from prograde to retrograde and
vice versa as seen from an inertial reference frame. Outside the bar
region,  chaos appears only in prograde orbits. Our results concerning
such connection are consistent and extend those obtained for midplane
orbits in the presence of only a spiral pattern in the  axisymmetric
Galactic model.

\end{abstract}


%\vfill\eject
%%%%%%%%%%%%%%%%%%%%%%%%%%%%%%%%%%%%%%%%%%%%%%%%%%%%%%%%%%%
\section{INTRODUCTION}\label{introd}

In the past few years it has been finally accepted the existence  of a
bar in the center of our Galaxy. Some studies providing evidence for
this Galactic component are the kinematical  data obtained in HI 21-cm
emission, CO, and CS (Sanders \& Prendergast 1974;  Liszt \& Burton
1980; Gerhard \& Vietri 1986; Binney \etal 1991), the  Galactic center
stellar distribution with Mira Variables from  IRAS (Harmon \& Gilmore
1988; Nakada \etal 1991; Weinberg 1992),  and the results of the
COBE/DIRBE satellite (Weiland  \etal 1994, and models based on these
observations, such as Dwek \etal 1995; Fux 1997; Freudenreich 1998;
Beaulieu \etal 2000;  Bissantz \etal 2003). Based on these evidences
and due to the expected importance of a non-axisymmetric  galactic
component to the stellar and gas dynamics, we have  constructed three
models for the gravitational  potential of the Galactic bar and
studied their dynamical effects on point masses orbiting the Galaxy.

Several authors have studied and modeled in many ways the
gravitational potential of bars. The simplest model is the
two-dimensional one, for which  the potential has the form
$\Phi_{Bar}(R,\varphi)=g(R)cos(2\varphi)$  (Contopoulos \&
Papayannopoulos 1980); $R,\varphi$ are polar coordinates  in the
Galactic plane, $\varphi$ being measured with respect to the long axis
of the bar; $g$ is the amplitude. Other models consider a
three-dimensional mass  distribution with similar stratification in
ellipsoids or prolate  spheroids given by

\begin{equation}  \rho_{Bar}(x,y,z) = \left\{ {\rho_c{(1-
m^2)}^n\atop 0}{,\ \ m\leq 1, \atop ,\ \ m\geq 1,}\right.
\label{rhob} \end{equation}

\noindent with $\rho_c$ the central density,  $m^2 =
\frac{x^2}{a^2}+\frac{y^2}{b^2}+\frac{z^2}{c^2}$, and  $a>b\geq c$ the
respective semi-axes (ellipsoid: $a>b>c$;  prolate spheroid: $a>b=c$).

\noindent With $n$ an integer, the density in equation (\ref{rhob})
corresponds to Ferrers ellipsoids (Ferrers 1877).  The ellipsoidal
case with $n=0$, i.e., an  homogeneous ellipsoid, has been considered
by Sanders \&  Tubbs (1980). Inhomogeneous cases ($n\neq 0$) with a
prolate  shape have been considered by Papayannopoulos \& Petrou
(1983);  Petrou \& Papayannopoulos (1983); Athanassoula \etal (1983);
Teuben \& Sanders (1985); Shlosman \& Heller (2002).  Inhomogeneous
ellipsoidal models are considered by  Pfenniger (1984), and Kaufmann
\& Contopoulos (1996).

A more elaborated model has been presented by Zhao (1996), who
employed the multipolar expansion technique  given by Hernquist \&
Ostriker (1992) to obtain the gravitational potential of a ``boxy''
mass distribution that is observed  in iso-density contours of
edge-on galaxies, and in our own Galaxy as well (Freudenreich 1998).

With the purpose of building a complete three-dimensional model for
the Milky Way, we have modeled the Galactic bar in three different
ways using the available observational parameters; the main
parameter being the observed density (Freudenreich 1998; see Section
\ref{models}),  that cannot be fitted with a simple model such as that
of equation  (\ref{rhob}). Our three models are based on the density
considered in Model S of Freudenreich (1998). The first model is an
inhomogeneous ellipsoid;  the second is an inhomogeneous prolate
spheroid, and the third one is a superposition of inhomogeneous
ellipsoids. The last model approximates the observed boxy-shaped
density stratification of the Galactic bar.

As an application of our models we have analyzed the structure of
Poincar\'e diagrams (or sufrace of section) corresponding to orbits in
the Galactic midplane and with the required energy to reach the inner
Galactic region.  This work extends a recent study in which the
structure was explored in a three-dimensional model for the spiral
arms (Pichardo \etal 2003, hereafter Paper I).  In both papers, the
axisymmetric background (Galactic) potential is that of Allen \&
Santill\'an (1991). A short description of the Galactic model was
given in Paper I.  The present paper shows also how the kinematics of
globular clusters could be altered by the presence of a bar, via
numerical integrations of the orbits  of six globular clusters in our
Galaxy. A more extended and detailed study of the effect of the
Galactic bar on the kinematics of the whole sample of globular
clusters with known absolute proper motions will be presented in a
future paper.

In Section \ref{fit_obs} we review  the   observational parameters of
the Galactic bar that are used in  our models. In Section \ref{models}
we describe the three models  for the Galactic bar. Section
\ref{results} gives our results:  the analysis of stellar midplane
orbits (Section \ref{orb_an}),  and the kinematics of six globular
clusters (Section \ref{cumulos}).  In Section \ref{conclusions} we
discuss the results and give our conclusions. Finally in the
Appendix, we give a detailed analytical description of our three
models and we include a force field analysis.

%%%%%%%%%%%%%%%%%%%%%%%%%%%%%%%%%%%%%%%%%%%%%%%%%%%%%%%%%%%

\section{OBSERVATIONAL PARAMETERS}\label{fit_obs}

In this Section we briefly discuss the observational  parameters
adopted in our models of the Galactic bar.

\noindent {\it The length of the bar}. This parameter (the length of
the bar) depends on another controversial parameter, namely the
position angle of the longest axis of the bar with respect  to the
line of sight. We have taken, mainly, models based  on the COBE/DIRBE
maps. In particular, for a position  angle $\varphi=20^{\rm o}$, these
models place the end of the bar at the galactocentric distance
$R_f=3.1-3.5$ kpc (Freudenreich 1998; Binney, Gerhard \& Spergel 1997;
Bissantz \& Gerhard 2002). Also based in the COBE/DIRBE data and a
Galactic model, for a given angular velocity of the bar, the
corotation resonance has been derived, giving $R_{cr} \sim 3.4$ kpc
(Bissantz, Englmaier, \& Gerhard 2003). This resonance gives a
constraint to the maximum length of the bar. These results are
as well consistent with observations of OH/IR stars  (Sevenster 1999)
and IRAS variables (Nikolaev \& Weinberg 1997).

\noindent {\it Axial ratios}. Parametric models based on  the
COBE/DIRBE data suggest axial ratios of approximately  10:3-4:3. Such
ratios are in reasonable agreement with  non-parametric models  of
Bissantz \& Gerhard (2002). In this work we   take the axial ratios
obtained in Model S of Freudenreich (1998).  These ratios are
10:3.76:2.59.

\noindent {\it Density law}. Although  exponential bars have been
frequently used and/or favored  by models and observations in other
galaxies (Lerner,  Sundin \& Thomasson 1999; Combes \& Elmegreen
1993),  in the case of the Galactic bar it appears that  toward its
center the density profile is not  exponential, but a flatter function
of the radius  (Alard 2001). We have adopted the  density law $\rho
\propto sech^2(R_S)$, with $R_S$ an effective radius proposed in Model
S of Freudenreich (1998) for the Galactic bar. Such model is based on
the COBE/DIRBE data set (1.25, 2.2,  3.5, 4.9 $\mu$m).

\noindent {\it The mass}. This is probably the most  difficult
Galactic parameter to determine, and consequently involves a large
uncertainty. Observations with the Space    Telescope  by Dwek \etal
(1995) allowed a photometric  determination of a mass for the Galactic
bulge-bar components  of $M_{Bar} \sim 1.3 \times 10^{10}$
M$\odot$. Matsumoto \etal (1982) and Kent (1992), using a dynamical
bulge modeled  as an oblate spheroid, determine theoretically a mass
$M_{Bar} \sim 1 \times 10^{10}$ M$\odot$. Zhao (1996),  through
observations of Galactic microlenses and  models restricted by
COBE/DIRBE data, finds a best model prediction of $M_{Bar} >2\times
10^{10}$ M$\odot$.  Weiner \& Sellwood (1999) employed a Ferrers bar
to perform hydrodynamical  simulations which reproduce the {\it l-v}
diagrams in the Galaxy, finding that only a massive  bar can produce
the huge non-circular motions observed  near the Galactic center. In
their best model they obtain a value $M_{Bar} \sim 9.8 \times 10^{9}$
M$\odot$ and  also determine the bulge mass,  $M_B \sim 5.4 \times
10^9$ M$\odot$.

\noindent {\it Angular velocity $\Omega_{Bar}$}. The angular  speed of
the bar is an important dynamical parameter still under debate.
Different models and observations  report a range of angular speeds;
in the last  decade, the reported values span the range  $40 \leq
\Omega_{Bar} \leq 70$ km s$^{-1}$ kpc$^{-1}$. Usual methods to
estimate this parameter are: 1) hydrodynamical simulations, which try
to reproduce the  {\it l-v} diagrams of  features such as the 3 kpc
arm (Englmaier \& Gerhard 1999;  Fux 1999; Weiner \& Sellwood 1999;
Bissantz, Englmaier, \& Gerhard 2003). 2) The method of ``orbital
resonances'', in which   features such as ring-like structures, or
observations of the local velocity distribution, are attributed or
related to orbital resonances assuming a Galactic model (for instance,
Dehnen 2000).  Given the position of the Lindblad inner resonance,
v.g., the model predicts an angular speed for the bar. 3) The ``direct
method'' (Tremaine \& Weinberg 1984),   a kinematic method that does
not rely upon a dynamical  model. This method utilizes estimations of
the surface brightness and measurements of the radial velocity along
the nodal line. It has been applied recently to our Galaxy by
Debattista, Gerhard \& Sevenster (2002)  using OH/IR stars.  We have
adopted, $\Omega_{Bar} = 60$ km s$^{-1}$ kpc$^{-1}$ (Bissantz,
Englmaier, \& Gerhard 2003). For a review of the observational
parameters relevant to the Galactic bar, see Gerhard (2002).

%%%%%%%%%%%%%%%%%%%%%%%%%%%%%%%%%%%%%%%%%%%%%%%%%%%%%%%%%%%

\section{GALACTIC BAR MODELS}\label{models}

In this section we make a brief introduction to our three  models of
the Galactic bar. For a detailed analytical description and an
analysis of the force fields, see Sections \ref{modelip}, \ref{mprol},
\ref{msuperp}, and \ref{analysis} in the Appendix. These models are
based on the density Model S of Freudenreich (1998,  hereafter MSF),
of which some properties have been given in Section \ref{fit_obs}. The
Model S has a density of the form $\rho \propto sech^2(R_S)$, with
$R_S$ given by,

\begin{equation} R_S = \left\{ \left[ \left( \frac {|x|}{a_x} \right) 
^{C_\perp}+ \left( \frac {|y|}{a_y} \right) ^{C_\perp} \right]
^{C_\parallel/C_\perp}+ \left( \frac {|z|}{a_z} \right) ^{C_\parallel}
\right\} ^{1/C_\parallel}.
\label{Rsfre}\end{equation}

With a Sun's galactocentric distance of 8.5 kpc, the scale lengths in
the directions $x, y, z$ (the major, middle, and minor semi-axes of
the bar lie along these directions respectively) are $a_x$ = 1.7 kpc,
$a_y$ = 0.64 kpc, $a_z$ = 0.44 kpc, and the exponents are
$C_\parallel$ = 3.5, $C_\perp$ = 1.57.  The effective boundary of  the
bar on the $x$-axis has a major semi-axis $a_{Bar}$ = 3.13 kpc,  which
sets the scaled distance $R_{end_S}$ = $a_{Bar}/a_x$ = 1.841.  In $R_S
\geq R_{end_S}$ the density has an additional Gaussian factor  with a
scale length $h_{end}$ = 0.46 kpc; this leads to a steep but  smooth
fall in density in the outer region.

Our first model for the Galactic bar is triaxial, as it seems to be
the general case of galactic bars; it is constructed with an ellipsoid
with  a {\it similar} mass distribution (Schmidt 1956) in order to
obtain the observed density law in the Galaxy: $\rho \propto
sech^2(R_S)$. In the same manner we have constructed a second model
with prolate shape that corresponds approximately to the Galactic bar
case (Section \ref{fit_obs}), with the same density stratification as
the ellipsoidal one. In both models we take $C_\parallel$ = $C_\perp$
= 2 in equation (\ref{Rsfre}), and equal scale lengths in the $y$ and
$z$ directions in the prolate case.  Our third model, constructed to
approximate the boxy mass distribution of MSF, is a superposition of
four ellipsoids  with the same density law as the ellipsoidal and
prolate models.

The gravitational potential of the modeled Galactic bar is easily
obtained from known results in potential theory (see, e.g. Mc Millan
1930; Kellogg 1953).

%%%%%%%%%%%%%%%%%%%%%%%%%%%%%%%%%%%%%%%%%%%%%%%%%%%%%%%%%%%%%%%%%%%%%%%%%%%%%

\section{RESULTS}\label{results}

\subsection{Orbital Analysis on the Galactic Plane}\label{orb_an}

As a first application of the models given in Section \ref{models}, we
make an analysis of orbital motion in the Galactic plane.  The orbital
analysis is made in the non-inertial reference frame attached to the
bar, labeled as the primed system of Cartesian coordinates
$(x',y',z')$. The $x'$-axis is taken as the line along the major axis
of the bar,  the $z'$-axis is perpendicular to the Galactic plane,
with its  positive sense toward the north Galactic pole, and the
$y'$-axis  is such that the $(x',y',z')$ axes form a right-handed
system.  The angular velocity of the Galactic bar, $\Omega_{Bar}$,
points  in the negative direction of the $z'$-axis, i.e., a clockwise
rotation as seen from the north Galactic pole. We take the mass of the
bar as $M_{Bar} = 9.8 \times 10^{9}$ M$\odot$, and its angular
velocity $\Omega_{Bar} =  60$ km s$^{-1}$ kpc$^{-1}$. The bar is
superimposed on a modified version of the axisymmetric Galactic 
potential of Allen \& Santill\'an (1991).

For motion in the Galactic plane, Jacobi's constant in the
non-inertial frame is

\begin{equation} E_J = \frac{1}{2}({v_x^\prime}^2 + {v_y^\prime}^2) + \Phi_{AS}(x^\prime,y^\prime) +
\Phi_{Bar}(x^\prime,y^\prime) -
\frac{1}{2}{\Omega^2_{Bar}}({x^\prime}^2 + {y^\prime}^2),
\end{equation}

\noindent with $\Phi_{AS}$ the axisymmetric Galactic potential of
Allen \&  Santill\'an (1991), and $\Phi_{Bar}$ the potential due to
the bar.  All the computations presented here were done using the
Bulirsch-Stoer algorithm of Press \etal (1992), conserving Jacobi's
constant within a relative  variation of
$\left|(E_{Ji}-E_{Jf})/E_{Ji}\right|\approx 10^{-11}$.

In Paper I we  made an analysis of orbital motion in the Galactic
plane using a three-dimensional model for the spiral arms. As in that
case, a convenient technique in our analysis is the use of Poincar\'e
diagrams, or surfaces of section, and periodic orbits.  In Paper I we
introduced the concept of {\it zero angular-momentum  separatrix},
loosely defined (but with a physical meaning discussed therein) as the
very narrow region in the surfaces of section that cleanly separates
prograde and retrograde orbits,  as seen from an inertial reference
frame. In that work we stress the point that such clean separation
will not appear in the customary surface of section diagrams which do not
take into account the information from the inertial frame. The reason
obeys to the fact that in the usual rotating frame, the definition of
prograde/retrograde is ambiguous, as an orbit may change its sense of
motion with time. 	 The separatrix is an exceptional set of
orbits: it consists of orbits with nearly zero angular momentum, for
which the  sign of the angular momentum  alternates  between positive
and negative {\it in the inertial frame}. The relevance of the
distinction between prograde and retrograde orbits is its apparent
connection with stochastic motion. For the spiral perturbation, we
found chaos only in prograde orbits, a region in the surface of section
diagram bounded by the separatrix.  In our application of this concept
to the Galactic spiral arms,  we noticed that when the mass of the
arms   was larger than a given value, the separatrix became wider. In
this section we analyze the behavior of this separatrix under the
gravitational potential of the Galactic bar.

Based on observational results, we have considered two experiments
with differences in the axisymmetric potential (specifically in the
central  component). Experiment I takes the axisymmetric potential
(halo, disk and bulge) with a bulge of approximately 45$\%$ of the
mass of the bar. Several authors consider that both structures, bulge
and bar, coexist (Norman, Sellwood \& Hasan 1996; Sevenster 1999;
Weiner \& Sellwood 1999; Zhao 2000).  In the experiment II the central
component (bulge) is completely removed,  leaving only the halo and disk
of the axisymmetric model, and the bar  component. With the purpose of
understanding the global effect of the  axisymmetric potential, we
have added one last experiment where we make  the analysis for the bar
alone, removing the axisymmetric component  (this will be called
experiment III).

\begin{figure}
\begin{center}
\epsscale{1.0}
\plotone{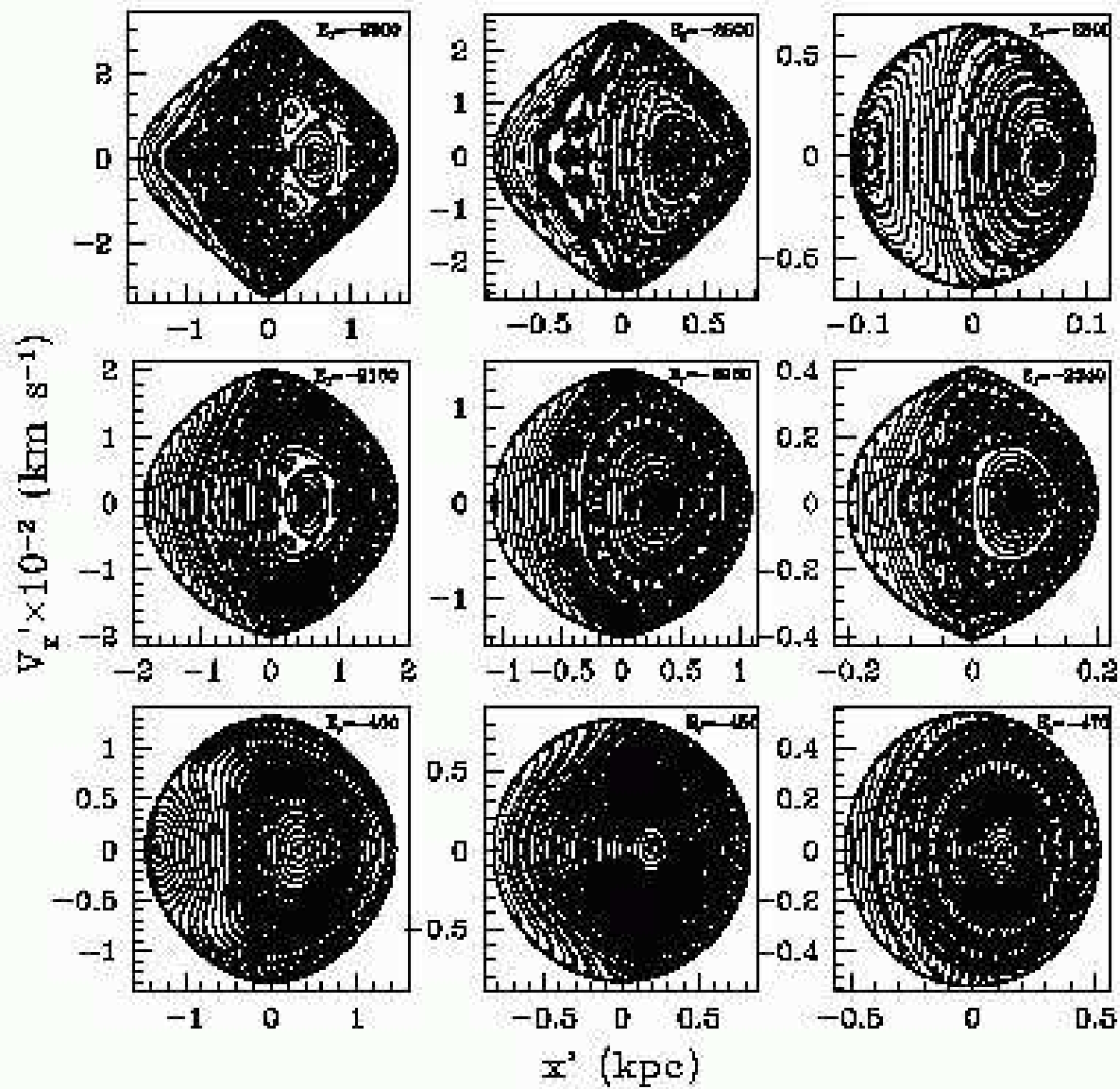}                      
\caption {Three examples of Poincar\'e diagrams using the 
ellipsoidal model, for experiments I, II and III, from top 
to bottom. In all cases the darkest regions represent
the orbits in the {\it separatrix}. The units 
of Jacobi's constant, $E_J$, are 100 km$^2$ s$^{-2}$ }
\label{fig.barelip}
\end{center}
\end{figure}

\begin{figure}
\begin{center}
\epsscale{1.0}
\plotone{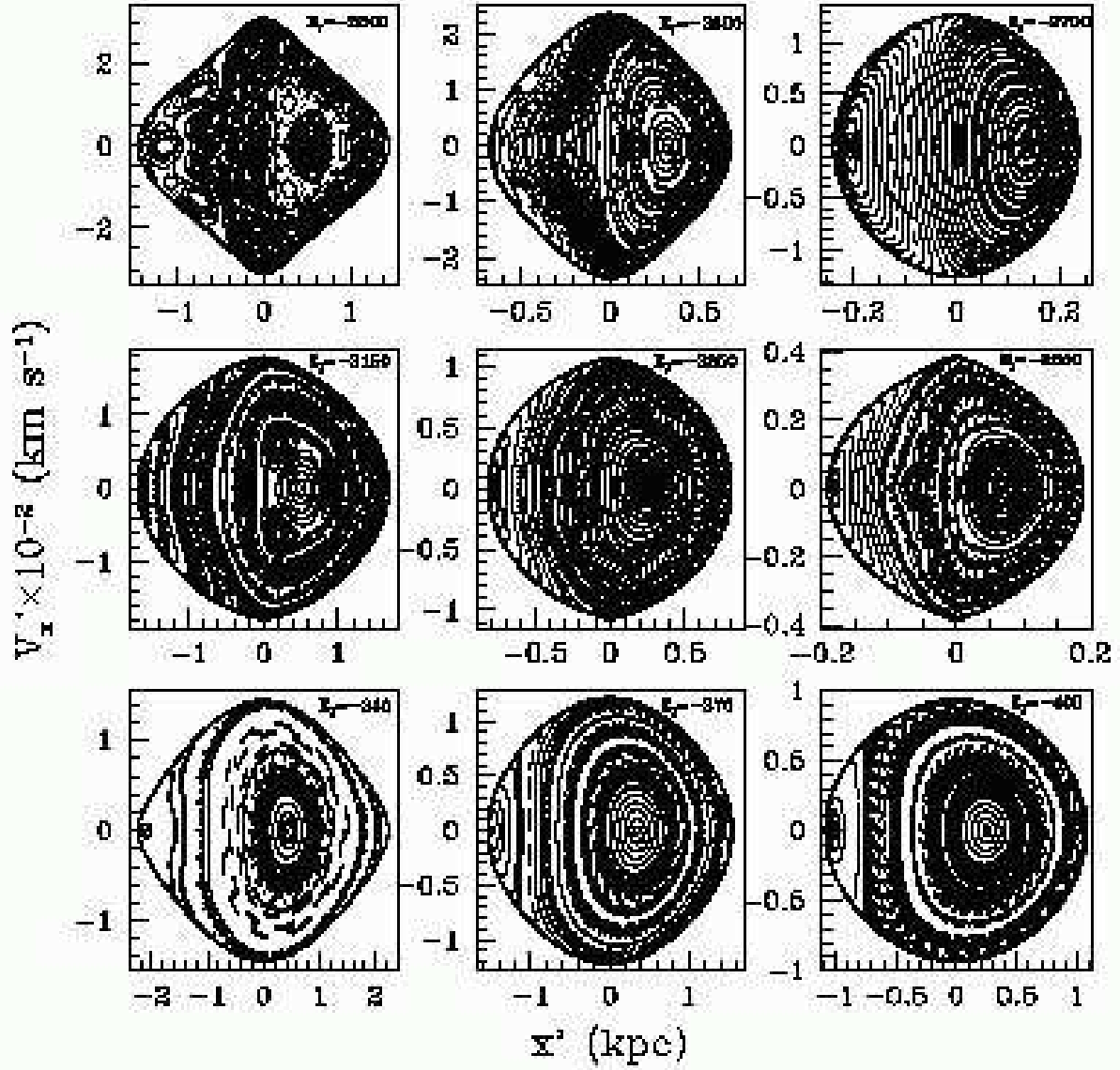}                      
\caption {Same as in Figure \ref{fig.barelip}, using the prolate 
model.}
\label{fig.barprol}
\end{center}
\end{figure}

\begin{figure}
\begin{center}
\epsscale{1.0}
\plotone{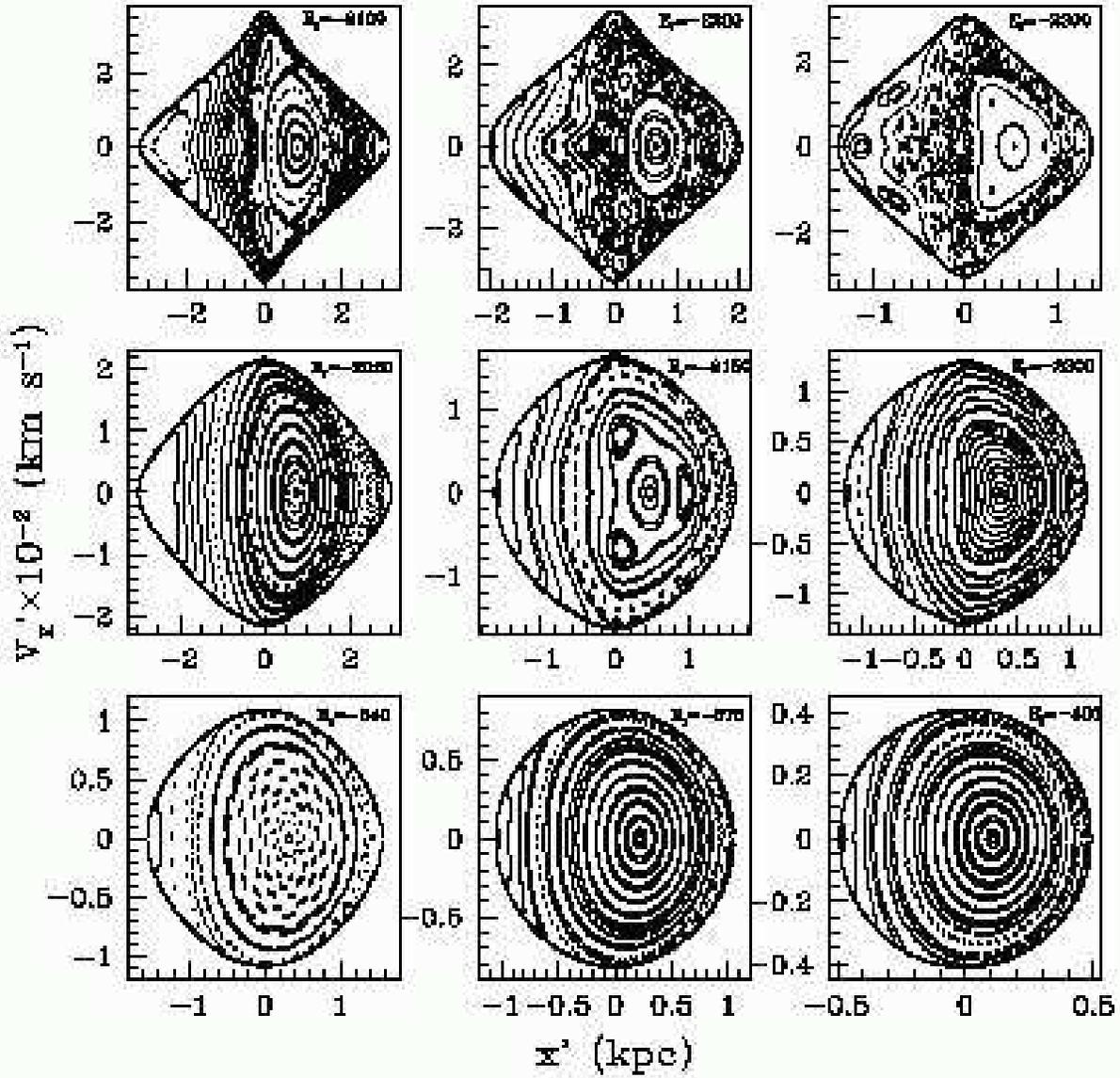}                      
\caption {Same as in Figure \ref{fig.barelip}, using the 
superposition model.}
\label{fig.barsuperp}
\end{center}
\end{figure}

Figure \ref{fig.barelip} shows nine Poincar\'e diagrams for orbits in
the presence of the ellipsoidal model of  the Galactic bar. Units of
Jacobi's constant ($E_J$ in the diagrams) are 100 km$^2$ s$^{-2}$.
The three diagrams on the top correspond to  experiment I, those in
the middle to  experiment II, and the bottom panels give the results
of  experiment III. In Figure \ref{fig.barprol} we present the same
experiments  as in Figure \ref{fig.barelip}, but now using the prolate
model.  Figure \ref{fig.barsuperp} shows the corresponding results
with the model of superposition of ellipsoids.

Comparing families of same $E_J$ for the three different models
(ellipsoidal, prolate and the superposition), we find very similar
results mainly between ellipsoidal and prolate bars in the 2-D orbital
dynamics case (for example the extension of the separatrix, retrograde
and prograde regions and the resonant families). Slight differences
are found with the superposition model.

\begin{figure}
\begin{center}
\epsscale{0.8}
\plotone{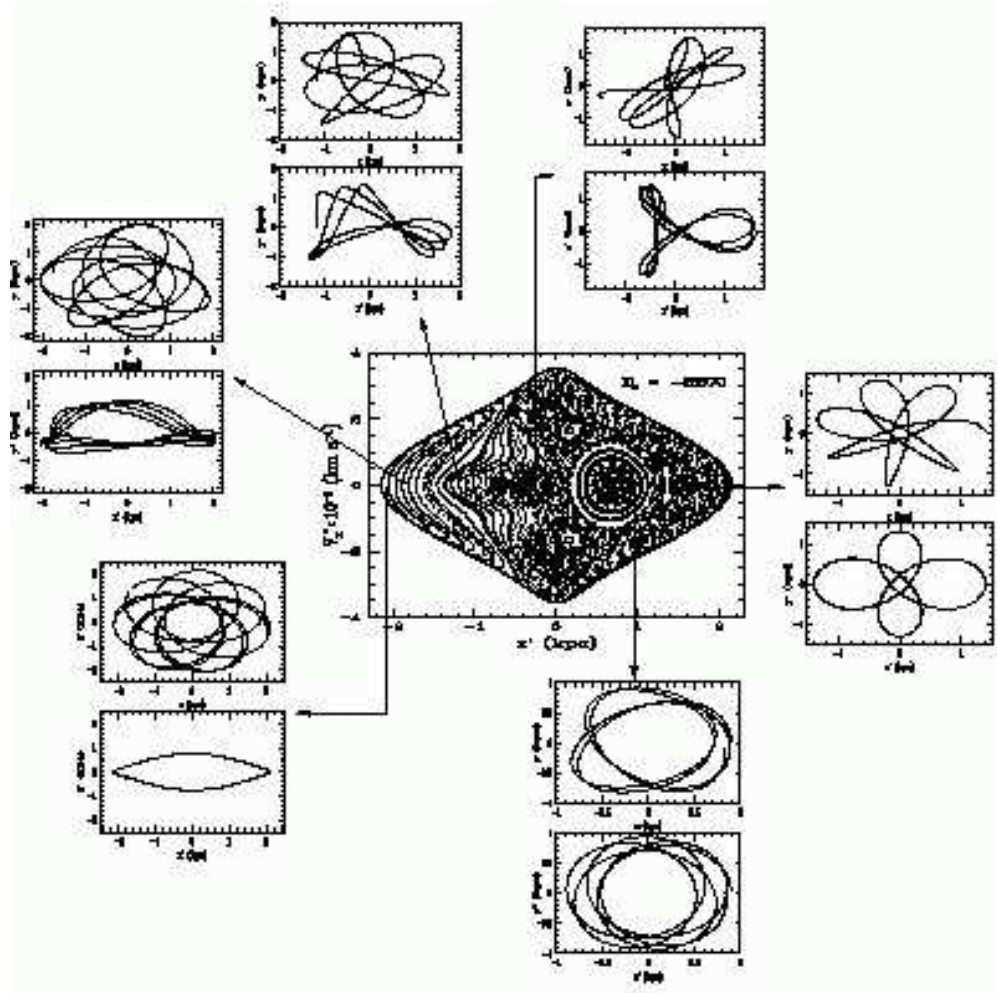}                      
\caption {Some examples of orbits in configuration space with the 
ellipsoidal model
for experiment I; the orbits are extracted 
from a given orbital family. Top panels of each frame correspond 
to the inertial reference frame, and bottom panels to 
the non-inertial frame that rotates with the bar.}
\label{fig.confighdbpb}
\end{center}
\end{figure}

\begin{figure}
\begin{center}
\epsscale{0.8}
\plotone{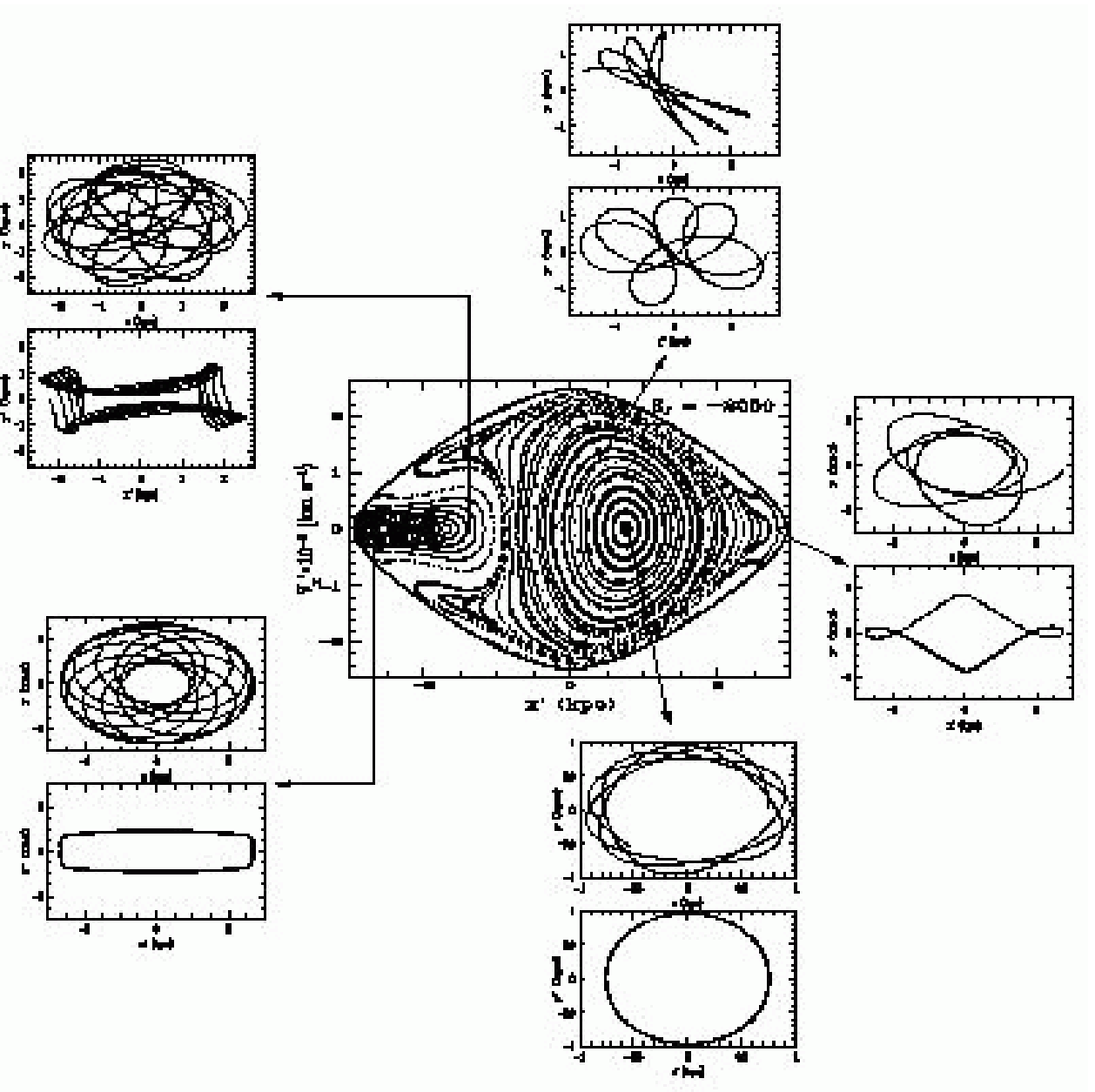}                      
\caption {Same as in Figure \ref{fig.confighdbpb}, for 
experiment II.}
\label{fig.confighdb}
\end{center}
\end{figure}

\begin{figure}
\begin{center}
\epsscale{0.8}
\plotone{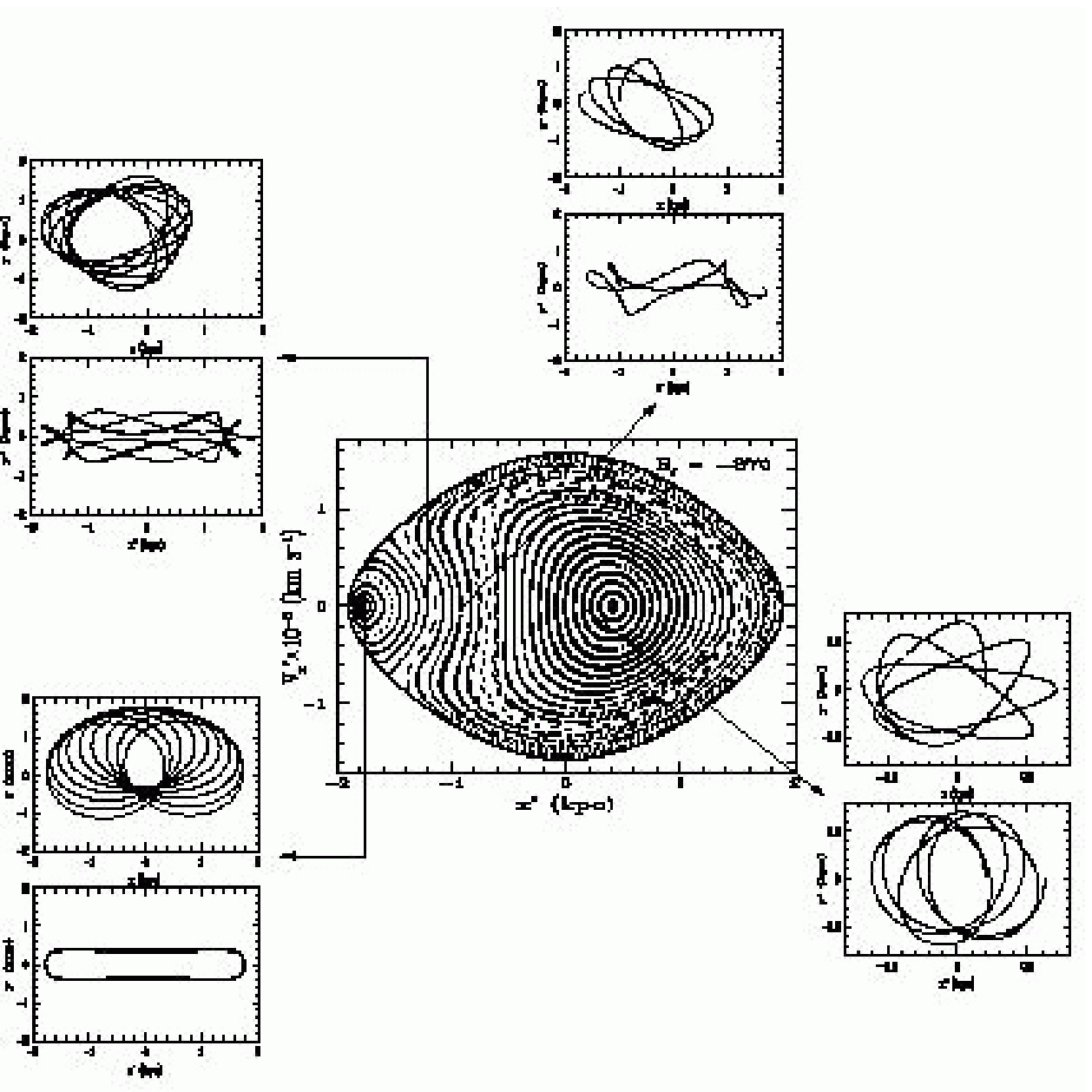}                     
\caption {Same as in Figure \ref{fig.confighdbpb}, for  experiment
III.}
\label{fig.configbs}
\end{center}
\end{figure}

Unlike our model with spiral arms (Paper I), where the separatrix was
defined by a very narrow curve in phase space, we notice that in all
our models with  the bar potential, the separatrix (which is shown in
all the phase space  diagrams  -Figures \ref{fig.barelip},
\ref{fig.barprol}, \ref{fig.barsuperp}, \ref{fig.confighdbpb},
\ref{fig.confighdb}, \ref{fig.configbs}- in darker points) is
considerably wider; i.e., we have more orbits that change their sense
of motion  from prograde to retrograde, and vice versa, {\it as seen
from an inertial  reference frame}. The width of this region depends
on the relative importance between the mass of the bar and the mass of
the central component (like a bulge). Orbits in this region might
result interesting, since orbital chaos seems to present a trend to
appear first in this type of orbits.

With the model of the ellipsoidal bar, in those cases that include a
central component, or bulge (experiment I),  for low values of $E_J$
(=-2800, for example)  the inner and most tied orbits are completely
ordered. For other values of $E_J$ (=-2300, for example), the onset of
chaos occurs in  orbits belonging to the separatrix. Beyond  the
corotation barrier (approximately located at 3.5 kpc from the
Galactic center), chaos dominates the prograde orbital region.  As in
the case of the spiral perturbation (Paper I),  there is no chaos in
the retrograde region (as seen from an  inertial reference frame).

To illustrate the behavior in real space of the orbits that compose
some Poincar\'e diagrams, we show in Figures \ref{fig.confighdbpb},
\ref{fig.confighdb}, and \ref{fig.configbs}, some examples of orbits
in the inertial (upper diagrams), and non-inertial (lower diagrams)
frames, with the ellipsoidal bar for experiments I, II, and III. In
these diagrams there are some examples of orbits like those that form
the prograde region (upper left panels in all cases). These orbits
have always a defined sense of motion in the same direction as the
bar. On other hand, in the non-inertial frame (bottom panel), orbits
have the characteristic changes of orbital sense and self-crossing
produced by accelerated reference frames. We also show some examples
of the periodic orbits $x_1$ (lower left panels) and $x_4$ (lower
middle panel in Figure \ref{fig.confighdbpb}). Orbits from the
separatrix are also shown (upper panels); these orbits are the ones
with the lowest angular momentum of each family $E_J$.

In the three models, the separatrix widens out going from experiment I
to III. This is due to the local effect of the  bar on the orbits;
i.e., stars feel locally stronger forces,  due to the bar, than the
axisymmetric force field that tends to force them to travel in orbits
with no change in the sense of rotation,  as seen from an inertial
reference frame. On these grounds,  we expect that, in galaxies where
the bulge (or any central  axisymmetric component) is very important
relative to the  bar (as in early type galaxies), orbits with a non
changing  sense of motion in the inertial system will dominate. On the
other hand, in galaxies where the central component is less important
relative to the non-axisymmetric component, orbits with a changing
sense of motion from prograde to retrograde and vice versa in the
inertial frame would be more important.

Regarding to chaos, Athanassoula (1990) finds that boxy-shape bars
produce more chaos in the orbital dynamics than elliptical bars,
however this model is two dimensional. This particular result is not
reproduced in our case, with a three dimensional bar potential, where
we find that chaotic regions in phase space are very similar in its
extension, and even in shape, to the results with the ellipsoidal and
prolate bars.

On the other hand, we notice, comparing experiments  I, II and III
(for any of our three bar models), that when the bulge is removed,
chaos disappears in orbits in the inner regions -- inside the extent
of the bar --. That is, the bar seems to be favored if there is no
central spherical component. The more massive is the bulge component,
the wider is the chaotic region in the inner region of the bar, until
this region covers all phase space, growing rapidly and destroying the
$x_1$ periodic orbits (this result is also obtained by Hasan \& Norman
1990, in a prolate spheroid, but the central component is in this case
a black hole). When the central spheric component is dominant, the bar
tends to disappear; in this case planar orbits would be fully
ordered again. Teuben \& Sanders (1985) find the same results in this
direction.

Chaos appears for some orbital families (the less tied up), and for
the parameters we have taken for the Galactic bar, it appears only in
the orbits that form the separatrix for the inner region of the bar,
and only in the prograde orbits for the outer regions of the bar,
crossing the corotation barrier. This last result (also found by Fux
2001, and Pfenniger 1984) is reproduced in all our models, unlike
Athanassoula \etal (1983) who find that chaos is not present in orbits
that reach corotation. Retrograde orbits do not present chaos in any
case (detailed results in these direction will be presented in a
future paper).

\begin{figure}
\begin{center}
\epsscale{0.4} \plotone{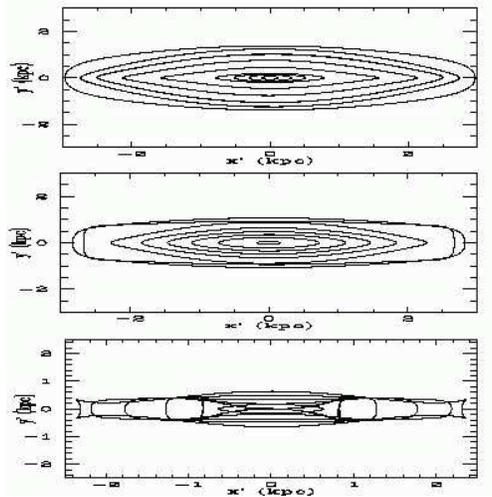}
\caption {Some examples of $x_1$ periodic orbits with the  ellipsoidal
model for experiments I (top panel), II (middle panel), III (bottom
panel)}
\label{fig.perhdbpb}
\end{center}
\end{figure}

The shape of the so called $x_1$ periodic  orbits seems to be related
with the real shape of bars (Teuben \& Sanders 1985).  We have
constructed some families of these kind of orbits for our ellipsoidal
bar. The shape of these sets of orbits  is also affected by the
presence of central spherical components (like a bulge). In this
manner, going from experiments I (top panel of  Figure
\ref{fig.perhdbpb}) to III (bottom panel of the same  Figure), orbits
change its shapes going from elliptical to ``boxy''  figures. The
higher the mass of the bulge or any central spherical component is,
the  rounder the periodic orbits are.

\subsection{Orbits of Some Globular Clusters}\label{cumulos}

Galactic orbits of globular clusters computed in an axisymmetric
Galactic potential have been studied by many authors (e.g., Allen \&
Martos 1988; Allen 1990; Brosche \etal 1991; Dauphole \etal 1996;
Dinescu, Girard, \& van Altena (1999, DGA99 hereafter); Brosche,
Odenkirchen, \& Geffert 1999). Properties of the computed orbits
(e.g., eccentricity,  peri- and apogalactic distances, maximum
distance from the Galactic plane, energy, z- component of the angular
momentum) have been usually related with the metallicity of the
cluster, inferring from this how the Galactic halo and disk were
formed.  However, the effect of both the spiral arms and the Galactic
bar remains obscure. In this section we compute the orbits of six
globular clusters in a Galactic potential which includes the Galactic
bar. We use the models for the gravitational potential of the bar
presented in Section \ref{models} (better described in the Appendix),
superimposed on an axisymmetric mass distribution obtained from the
axisymmetric Galactic model of Allen \& Santill\'an (1991).  In a more
detailed study underway, we analyze the Galactic orbits of the 38
globular clusters with known revised absolute proper motions compiled
by DGA99, considering the Galactic bar and the spiral arms.

As in Section \ref{orb_an}, we take the mass of the bar and its
angular velocity as $M_{Bar} = 9.8 \times 10^9$ M$\odot$,
$\Omega_{Bar} =  60$ km s$^{-1}$ kpc$^{-1}$.  The mass of the bulge in
the Galactic model of Allen \& Santill\'an (1991), $M_B = 1.4 \times
10^{10}$ M$\odot$, is reduced to $M_B = 4.26 \times 10^9$ M$\odot$ to
account for the added bar component (thus assuming this bar is part of
the bulge).

At time $t = 0$ the major axis of the bar makes an angle of $\sim
20^{\rm o}$ with the Sun-Galactic center line (Freudenreich 1998). We
integrate the orbits backward in time during $\sim 1.5 \times 10^{10}$
yr, using the Bulirsch-Stoer algorithm of Press \etal (1992). The
orbits are computed in the non-inertial reference frame of the bar
(see Section \ref{orb_an}), where Jacobi's constant can be used to
check the numerical integration. In each orbit we localize all the
points where the distance to the Galactic center, $r$, has a local
maximum or minimum, and the points where the absolute value of the
$z$- coordinate (perpendicular to the Galactic plane), $|z|$, has a
local maximum. An eccentricity $e = (r_{max} - r_{min})/(r_{max} +
r_{min})$ is computed with successive values of extrema in $r$. Also,
we compute in the Galactic inertial frame the values of the energy per
unit mass, $E$, and the $z$- component of the angular momentum per
unit mass, $h$, at all orbital points with an extremum in $r$ or
$|z|$.  Thus, with $E,h$ at these points we sample the variation of
these important  quantities, which are otherwise constant when using
an axisymmetric Galactic model.

From the 38 globular clusters listed by DGA99, the six clusters we
have chosen are among the clusters that according to their Table 5
have perigalactic distances lying in the region of the Galactic bar,
as  computed in their axisymmetric Galactic model. Thus, we consider a
sample  of clusters for which the effect of the Galactic bar is
expected to be  important. The six globular clusters are: NGC 5139
($\omega$ Cen), NGC 6093 (M 80), NGC 6144, NGC 6171 (M 107), NGC 6218
(M 12), and NGC 6712.  Table 2 of DGA99 lists the data from which the
initial conditions of the orbits can be obtained. The adopted solar
motion is $(U,V,W)$ = (-11.0, 14.0, -7.5) km s$^{-1}$ ($U$ positive
outward from the Galactic center), 220 km s$^{-1}$ the rotation
velocity of the LSR, and 8.5 kpc the Galactocentric distance of the
Sun.  We have taken into account a misprint in the radial velocity of
NGC 6218 listed in Table 2 of DGA99; this  velocity is negative,
according to Pryor \& Meylan (1993). Also, a minus  sign is missing in
the $U$- velocity with respect to the LSR listed for NGC 6144.

In Table 1 we give the orbital properties of the six globular
clusters,  computed in the axisymmetric Galactic model of Allen \&
Santill\'an (1991)  during an interval of 1.5 $\times 10^{10}$ yr
backward in time. Average  values are given for $r_{min}$, $r_{max}$,
$|z|_{max}$, $e$, and in the last two columns the values of the
constants $E,h$. Comparable results are listed in Table 5 of DGA99,
obtained with their axisymmetric Galactic model.

\begin{figure}
\begin{center}
\epsscale{1.0} \plotone{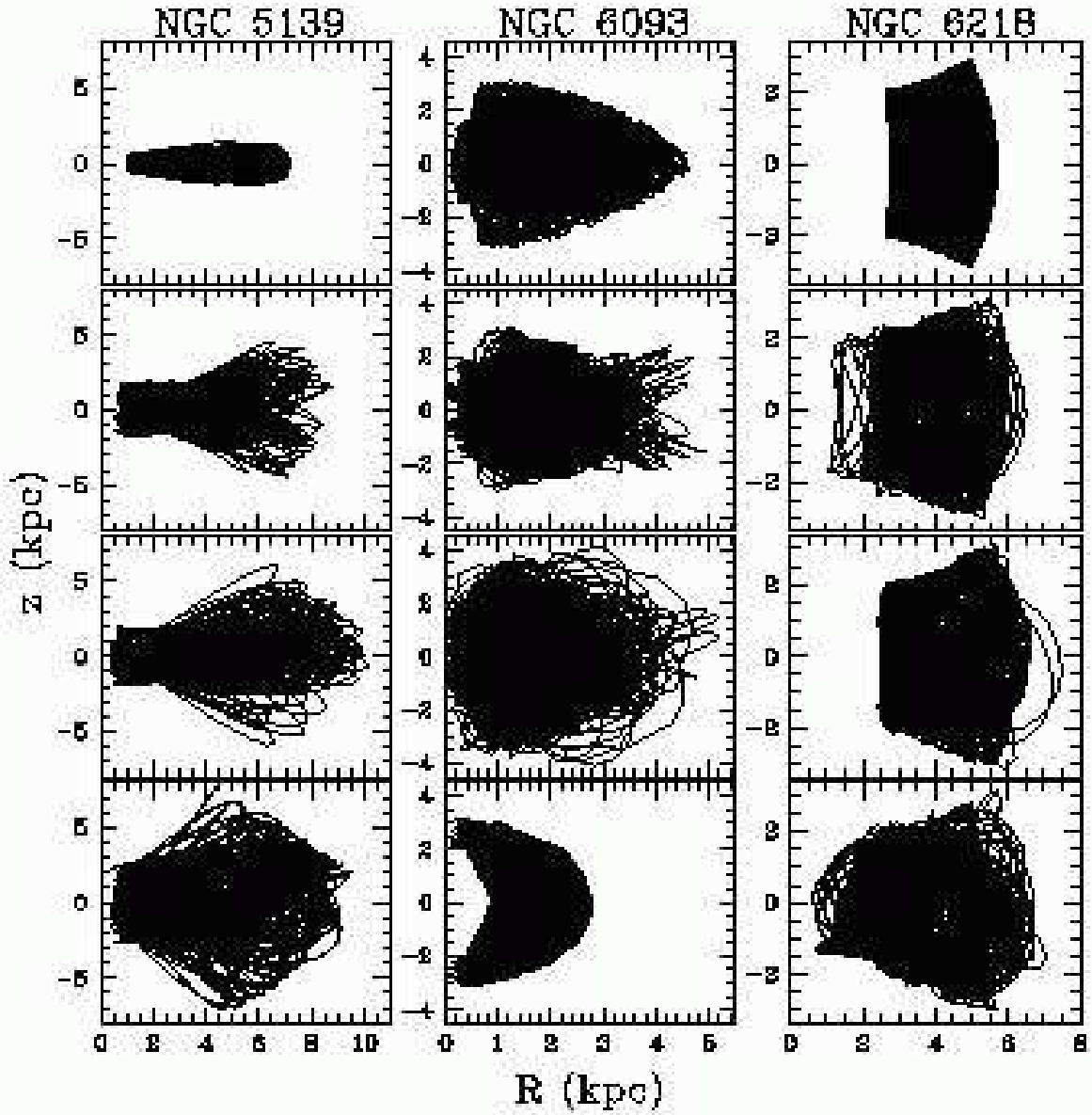}
\caption {Meridional orbits of NGC 5139 (left column), NGC 6093
(middle column), and NGC 6218 (right column). The orbits in the upper
panels are computed with the axisymmetric Galactic model of Allen \&
Santill\'an (1991). The second, third, and fourth rows give the orbits
in a Galactic model using the ellipsoidal, prolate, and superposition
models of the Galactic bar, respectively.}
\label{fig.orbs}
\end{center}
\end{figure}

The orbital properties of the six clusters obtained with the three
models of the Galactic bar are presented in Table 2. Minimum and
maximum values of $E$ and $h$ are given in the last four columns. The
three lines in each entry correspond to the ellipsoidal, prolate, and
superposition models, in this order. The average values in this table
are computed over the 1.5 $\times 10^{10}$ yr time interval.  Negative
values of $h$ mean the orbital motion is retrograde, i.e., in the
opposite sense to the actual rotation of the bar (or in the same sense
of the backward-in-time rotation of the bar). The averages in Table 2
are approximately similar in the three models of the Galactic bar,
except in the case of NGC 6093, in which the superposition model gives
an orbital evolution quite different from that obtained with the
ellipsoidal and prolate models. To illustrate the dependence of the
detailed orbital evolution on the model, Figure \ref{fig.orbs} shows
the meridional orbits of NGC 5139, NGC 6093, and NGC 6218. The upper
panels give the orbits in the axisymmetric Galactic model of Allen \&
Santill\'an (1991), the panels in the second row correspond to the
ellipsoidal model, those in the third row correspond to the prolate
model, and the panels at the bottom give the orbits in the
superposition model. Notice the different scales used in the figure;
these scales are the same in a given cluster (except for the orbit of
NGC 6218, where the scales are not the same in the vertical and
horizontal axes). Comparing with the orbital evolution obtained with
the axisymmetric model, this figure shows that  the inclusion of the
bar can increase the apogalactic distance and the $z$- distance from
the Galactic plane, and also decrease in some times the perigalactic
distance. Cases like NGC 6093 computed in the superposition model
deserve a more detailed analysis.

\begin{figure}
\begin{center}
\epsscale{1.0} \plotone{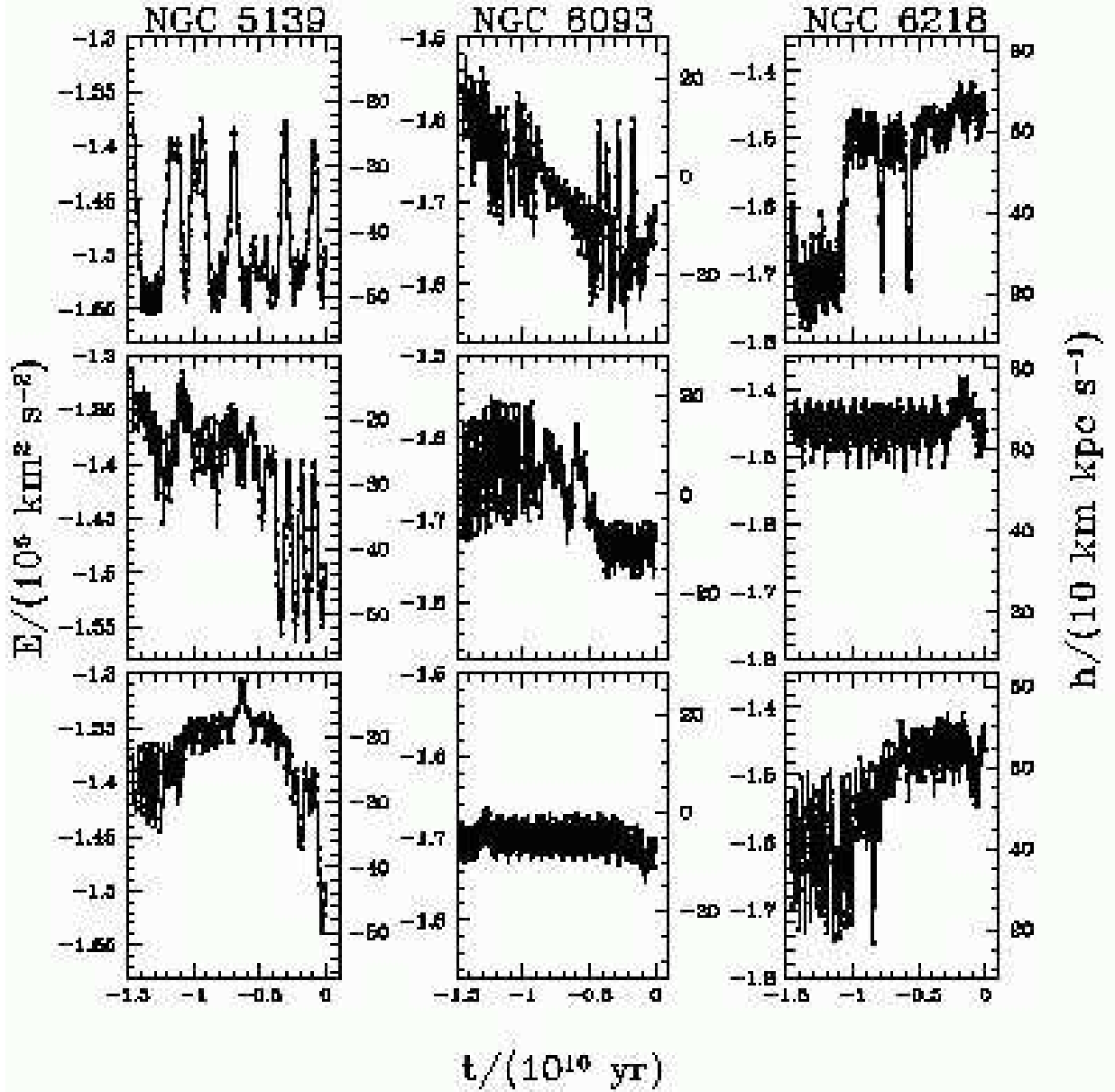}
\caption {Variations of $E$ and $h$ in the orbits of NGC 5139, NGC
6093, and NGC 6218. Upper panels: ellipsoidal model of the bar; middle
panels: prolate model; bottom panels: superposition model. $E$ is read
on the left scale, $h$ on the right scale.}
\label{fig.Eh}
\end{center}
\end{figure}

In Figure \ref{fig.Eh} we give the details of the variations of $E$
and $h$ in the orbits of the clusters in Figure \ref{fig.orbs}. In
each panel, $E$ is read on the left scale and $h$ on the right
scale. The upper, middle, and bottom panels correspond to the
ellipsoidal, prolate, and superposition models of the bar,
respectively. Figure \ref{fig.Eh} shows an interesting additional
effect which can be produced with the bar: large variations in $E$ and
$h$ with the possibility of a change in the sense of orbital rotation,
as measured in the Galactic inertial frame. This effect appears in the
orbit of NGC 6093 computed with the ellipsoidal and prolate models of
the bar; it barely appears at some time with the superposition model
(see value of $h_{max}$ in Table 2). We are analyzing this effect
particularly in the clusters reaching the inner Galactic region. This
effect might account for the observed retrograde motions in some
clusters.

\section{CONCLUSIONS}\label{conclusions}

We present three models for the gravitational potential of the
Galactic bar, based on the mass distribution for this component given
by Model S of Freudenreich (1998).  These models are an ellipsoid, a
prolate spheroid, and a superposition of four ellipsoids. The models
can be easily implemented for a numerical integration of orbits in a
non-axisymmetric Galactic potential. In particular, our third model,
which is a superposition of four inhomogeneous ellipsoids, gives a
good approximation to the boxy mass distribution of the Galactic
bar. Thus, in this model the resulting gravitational potential might
give relevant results in the analysis of orbits reaching the region of
the bar. For orbits lying outside the bar, the detailed modeling of
the shape of the bar is less important, and any of the three models
can be used.

We have applied our models to orbits in the Galactic plane in the
inner Galactic region, and to orbits of some globular clusters. We
find that the bar produces a dispersion on the energy and angular
momentum, as measured in the Galactic inertial frame. In particular,
for orbits with the $z$- component of angular momentum close to zero,
this dispersion effect can make an orbit oscillate between prograde
and retrograde, resulting in a wider separatrix. In the case of
globular clusters, the bar might be responsible for the observed
orbital prograde-retrograde distribution. In general, the relative
importance of  the bar with respect to a central axisymmetric
component determines the  dominant stellar sense of motion, i.e., the
larger is this ratio  $(M_{Bar}/M_{Cen}$, where $M_{Bar}$ is the bar
mass and $M_{Cen}$ is the mass of  a central component like a bulge),
the larger is the population of stars  that will change their sense of
motion from prograde to retrograde  (and vice versa), as seen from an
inertial reference frame.

In a preliminary analysis of chaotic regions, we have found that a
central axisymmetric component induces the onset of orbital chaos in
the inner region of the bar, and chaos mainly appears in the orbits
that change their sense of motion in the inertial frame of reference,
i.e. those that form the separatrix. Outside the bar, chaos only
appears after the corotation barrier and only in the prograde
orbits. As the central component mass is reduced or disappeared, chaos
diminishes or is completely removed for orbits in the inner region of
the bar. It is remarkable that the connection between stochastic
motion and sense of motion measured from the inertial frame, found for
the relatively weak spiral perturbation, is preserved under the much
stronger perturbation of the bar. However, the properties of the
separatrix are still a subject we consider deserves a more detailed
study.

For globular clusters, chaos is found (v.g., Allen and Martos 1988) in
the axisymmetrical potential with no need of a perturbation such as
the presence of a spiral pattern or a bar. Chaos in those systems is
seemingly related to the impulsive nature of the rapid passing of the
cluster through the large mass concentration at the central regions of
the Galaxy. For midplane motion, results concerning the connection
between the angular momentum and chaos are apparently indicating that
the physical agent has to do with a secular effect; i.e., prograde
orbits with respect to the general motion of the perturbing mass,
spiral or bar, tell us about longer times under their influence than
that from a rapid encounter, as that expected from retrograde
motion. A picture able to include a general explanation for both
mechanisms triggering chaotic motion with a physical flavor seems
necessary. In the case of planar orbits, Paper I invoked the
overlapping of resonances as the standard explanation for the
different phenomenology between prograde and retrograde motion in
regard to stochasticity. In three dimensional motion, resonances
involving vertical oscillations could be the lacking piece for a
unified scheme.

\section*{Acknowledgments}

This work was partially supported by Universidad Nacional Aut\'onoma
de M\'exico (UNAM) under DGAPA-PAPIIT grant IN114001, and CONACYT grant
36566-E. Calculations were carried out using the Origin Silicon Graphics
supercomputer of DGSCA-UNAM. We thank our anonymous referee for helpful 
suggestions.
\clearpage

%%%%%%%%%%%%%%%%%%%%%%%%%%%%%%%%%%%%%%%%%%%%%%%%%%%%%%%%%%%%%%%%%%%%%%%
\appendix
\begin{appendix}
\centerline{\bf APPENDIX}
In this appendix we present the detailed analytic derivation 
of the potential models for the Galactic bar proposed in this 
work (briefly described in Section \ref{models}).

\section{The Ellipsoidal Model}\label{modelip}
Our first model for the Galactic bar is an 
inhomogeneous ellipsoid. Its similar mass distribution has a density,

\begin{equation}\hskip -3.1cm\rho(R_S)=\rho_0 sech^2(R_S), \ \ \ 
R_S\leq R_{end_S} \atop \rho(R_S)=\rho_0 sech^2(R_S) 
e^{-(R_S-R_{end_S})^2/h_{end_S}^2}, \ \ \ R_S\geq R_{end_S}
\label{denel}\end{equation}

\noindent with $R_S = \left\{\frac{x^2}{a_x^2}+\frac{y^2}{a_y^2}+
\frac{z^2}{a_z^2}\right\}^{1/2}$ , $h_{end_S}= h_{end}/a_x$.  The
second line in equation (\ref{denel}) shows the Gaussian factor.
Equation (\ref{denel}) corresponds to equations (13) and (14) of
Freudenreich (1998),  but we have corrected a misprint in his Gaussian
factor, adapting this as given above.

\begin{figure}
\begin{center}
\epsscale{1.0}
\plotone{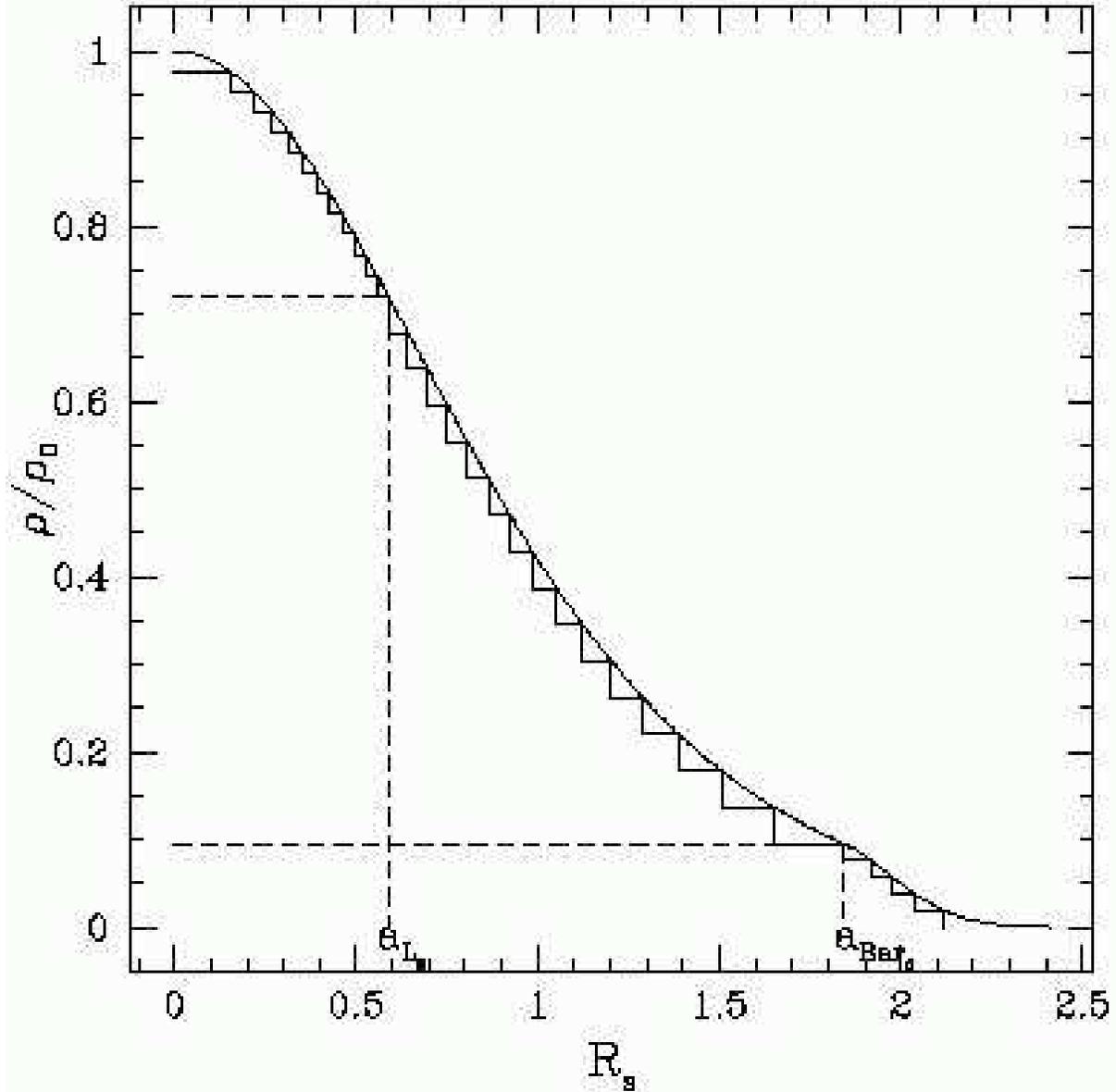}                      
\caption {Superposition of homogeneous ellipsoidal components  to
approximate the density law of the Galactic bar. In the vertical axis
the three intervals in density are shown. In this example, $a_L$ = 1
kpc, $N_1$ = 12, $N_2$ = 15, $N_3$ = 5.\ $a_{L_S}$ and $a_{Bar_S}$ are
the scaled $a_{L}$ and $a_{Bar}$ -i.e. divided by the scale length-.}
\label{fig.parti}
\end{center}
\end{figure}

As in the case of the construction of an inhomogeneous spheroid with
{\it similar} strata  using homogeneous spheroidal components (Schmidt
1956), our ellipsoidal model for the bar approximates the density in
equation (\ref{denel}) with a step-stair function. Figure
\ref{fig.parti} shows the density function normalized by the central
density, $\rho/\rho_0$, vs. $R_S$. Each stair step represents an
homogeneous ellipsoidal  component. By taking a large number of these
components the     accuracy of the approximation improves rapidly.

Since the density function does not have a constant gradient, we have
taken three intervals or regions in the scaled distance $R_S$ to
specify different number of components according to the dominant
gradient of the density function  (see Fig. \ref{fig.parti}). In the
inner region the density interval is divided in $N_1$ subintervals,
giving $N_1-1$ homogeneous ellipsoidal components, each one with a
density $\Delta\rho_1$.  The middle and outer regions have,
respectively, $N_2, N_3$ subintervals and  components, with
corresponding densities $\Delta\rho_2, \Delta\rho_3$.  The inner
region ends at a major semi-axis $a_L$, $0 < a_L < a_{Bar}$, i.e., at
$R_S = a_{L_S} = a_L/a_x$; the middle region ends at the effective
major semi-axis of the bar $R_{end} \equiv a_{Bar}$ = 3.13 kpc, i.e.,
at $R_S = a_{Bar}/a_x \equiv a_{Bar_S}$ = 1.841 .  The outer region
contains the Gaussian factor in its density. Figure \ref{fig.parti}
shows an example with $N_1$ = 12, $N_2$ = 15, $N_3$ = 5, and $a_L$ =
1.0 kpc.

This procedure of taking three regions to account for different
gradients in density will not have relevance in the limit of large
partition numbers ($N_1,N_2,N_3$), which is ultimately the limit to
consider in this formulism. This analysis will be necessary in our
three models; in Section \ref{analysis} we give a comparison  of the
total force fields of the bar obtained with different partitions.

The densities of the ellipsoidal components in the three regions are

\begin{eqnarray}\Delta\rho_1&=&\frac{\rho_0}{N_1}\left\{1-
sech^2(a_{L_S})\right\}\cr \Delta\rho_2
&=&\frac{\rho_0}{N_2}\left\{sech^2(a_{L_S})-sech^2(a_{Bar_S})\right\}
\label{dens}\end{eqnarray}
\vskip -0.9cm
\[\hskip -2.9cm \Delta\rho_3 \ \ =\ \frac{\rho_0}{N_3}sech^2(a_{Bar_S}),\]

\noindent then, with equation (\ref{denel}) the scaled dimensions
$R_S$ of these components in the inner and middle regions are

\begin{equation} R_{s_i} = sech^{-1}\left\{ 1-\frac {i}{N_1}
\left[ 1-sech^2(a_{L_S})\right] \right\}^{1/2}, \ \ i=1,2,...,N_1-1
\end{equation}

\begin{equation} R_{s_{N_1+j}} = sech^{-1}\left\{ sech^2(a_{L_S})-
\frac {j}{N_2}\left[ sech^2(a_{L_S})-sech^2(a_{Bar_S})\right]
\right\}^{1/2}, \ \ j=0,1,2,...,N_2-1 \end{equation}

\noindent
In the outer region the scaled dimensions are obtained solving the
equation

\begin{equation} \left(1-\frac{k}{N_3}\right) sech^2(a_{Bar_S})- 
sech^2R_{s_{N_1+N_2+k}} e^{-\left(R_{s_{N_1+N_2+k}}-a_{Bar_S}
\right)^2/h_{end_S}^2} = 0, \ \ k=0,1,...,N_3-1 \end{equation}

Defining $\zeta \equiv a_y/a_x$ and $\xi \equiv a_z/a_x$, the volume
of an ellipsoidal component with a scaled major semi-axis $a_{s_l} =
R_{s_l}, l=1,2,...,N_1+N_2+N_3-1$, is $V_l = \frac {4}{3}\pi\zeta\xi
a_l^3$, with $a_l = a_{s_l}a_x$. Thus, with $\Delta_1 = \Delta\rho_1/
\rho_0$, $\Delta_2 = \Delta\rho_2/\rho_0$, and $\Delta_3 =
\Delta\rho_3/\rho_0$ (these quantities obtained in terms of $N_1, N_2,
N_3$ from equation (\ref{dens}))  the total mass of the ellipsoidal
components (= mass of the bar) is

\begin{equation} M_{Bar} = \frac{4}{3} \pi \zeta \xi \rho_0 
\left\{ \Delta_1 \sum_{i=1}^{N_1-1} a_i^3 + \Delta_2
\sum_{j=0}^{N_2-1} a_{N_1+j}^3 + \Delta_3 \sum_{k=0}^{N_3-1}
a_{N_1+N_2+k}^3 \right\}.
\label{melip} \end{equation} 

For a given total mass $M_{Bar}$, this last equation gives the
corresponding central density $\rho_0$; thus the densities of the
components follow from equation (\ref{dens}). Finally, with the
densities and dimensions of these components, their gravitational
potentials are obtained with standard potential theory  (e.g., Kellogg
1953). The potential of the bar at a given point  in space is the sum
of these potentials at this point.

\section{The Prolate Model}\label{mprol}

Even though it is believed that bars in general are triaxial
structures, there are strong pieces of evidence pointing to an
approximately prolate Galactic bar (Freudenreich 1998). Therefore it
is appropriate to consider a prolate shape. A model of this type for
the Galactic bar  is analytically simpler than the ellipsoidal one,
and it makes an orbital integration considerably faster.

The density in this model has again the form given in equation
(\ref{denel}), but now $R_S = \left\{\frac{x^2}{a_{x_p}^2} +
\frac{y^2+z^2}{a_{y_p}^2}\right\}^{1/2}$, with $a_{x_p} = a_x$,
$a_{y_p} = \frac {1}{2}(a_y+a_z)$.

Schmidt (1956) gives the procedure to obtain the gravitational
potential and force due to an oblate spheroid with a {\it similar}
mass distribution $\rho(a)$, where $a$ is the major semi-axis of a
{\it similar} oblate surface. Following his procedure one can readily
obtain the corresponding expressions for a prolate spheroid with the
same type of mass distribution. With the $x$-axis being the long axis
of the prolate spheroid, at a given point ${\bf r}=(x,y,z)$ the
acceleration components along and perpendicular to this  axis are (in
units in which the Universal gravitational constant G = 1)
 
\begin{equation} -\frac{\partial\Phi}{\partial x} =
-4\pi\ e_p^{-3}\ (1-e_p^2)\ x\ \int^{\beta_p'}_0 \rho(a(\beta_p))
\frac{sin^2 \beta_p}{cos\beta_p} d\beta_p,
\label{dpotx} \end{equation}

\begin{equation} -\frac{\partial\Phi}{\partial R} =
-4\pi\ e_p^{-3}\ (1-e_p^2)\ R\ \int^{\beta_p'}_0 \rho(a(\beta_p))
\frac{sin^2 \beta_p}{cos^3\beta_p} d\beta_p,
\label{dpotr} \end{equation}

\noindent with $e_p = (1-(\frac {b_{sph}}{a_{sph}})^2)^{1/2}$ the
eccentricity of the spheroid, $a_{sph}, b_{sph}$ its major and minor
semi-axes, $R = (y^2 + z^2)^{1/2}$, $\beta'_p = sin^{-1}\ e_p$ if
${\bf r}$ is internal to the spheroid, and $\beta'_p$ the solution of
$x^2 sin^2\beta_p' + R^2 \tan^2 \beta_p' = a^2_{sph} e_p^2$ if ${\bf
r}$ is an external point. The function $a(\beta_p)$ is obtained from
$x^2 sin^2\beta_p + R^2 \tan^2 \beta_p = a^2 e_p^2$.

The potential at an internal point ${\bf r}$ is

\begin{equation} \Phi = -4\pi e_p^{-1} (1-e_p^2)\left
\{\int_0^{a({\bf r})} \rho(a) a \ln \frac{1+sin\ \beta_p}{cos\beta_p}
da + \frac{1}{2} \ln \frac{1+e_p}{1-e_p} \int_{a({\bf r})}^{a_{sph}}
\rho(a) a~da \right\},
\label{potpi} \end{equation}

\noindent with $a({\bf r}) = \left(x^2 +
\frac{R^2}{1-e_p^2}\right)^{1/2}$ the major semi-axis of the {\it
similar} spheroidal surface passing through the point ${\bf r}$. If
${\bf r}$ is an external point the potential is

\begin{equation} \Phi = -4\pi\ e_p^{-1}\ (1-e_p^2)\ \int_0^{a_{sph}}
\rho(a) a \ln \frac{1+sin\ \beta_p}{cos\beta_p} da.
\label{potpe} \end{equation}

Analytical solutions of equations (\ref{dpotx}) - (\ref{potpe}) are
difficult to obtain   for the density of the form of  equation
(\ref{denel}) applied to a prolate bar (notice that  equation
(\ref{denel}) gives the density at the scaled distance $R_S$; in
equations above we need the density at the $unscaled$ semi-axis
variable $a = R_Sa_{x_p} \equiv R_Sa_x$ along the major axis). Thus,
we need again an approximation for the density function. Due to the
fact that prolate spheroids are mathematically simpler than
ellipsoids, we can make a step further  in the representation of the
density. Instead of a step-stair  representation, as in the
ellipsoidal model, a better approximation  is a set of linear
functions, a polygon. With this approximation, the  analytical
solution of equations (\ref{dpotx}) - (\ref{potpe}) is  readily
obtained for each linear part.

We use the same procedure of assigning a partition in scaled distance
$R_S$ as in the ellipsoidal model. First, we re-number the scaled
distances $R_{s_l}, l = 1,2,...,N_1+N_2+N_3-1$, taking $n = l+1$ and
$R_{s_1} = 0$. The densities at these re-numbered scaled distances
$R_{s_n}$ are $\rho(R_{s_n}), n=1,2,...,N_1+N_2+N_3$. Then, with

\begin{equation} p_{0_n} = \frac {R_{s_{n+1}}\rho(R_{s_n})-R_{s_n}
\rho(R_{s_{n+1}})}{R_{s_{n+1}}-R_{s_n}}, \label{p0n} \end{equation}

\begin{equation} p_{1_n} = \frac{\rho(R_{s_{n+1}})-\rho(R_{s_n})}{R_{s_{n+1}}-
R_{s_n}}, \label{p1n} \end{equation}

\noindent the set of linear functions which approximate the density is

\begin{equation} \rho_n^\star[R_S] = p_{0_n} + p_{1_n} R_S,\ \ R_{s_n} \leq R_S
\leq R_{s_{n+1}}, \ n=1,2,...,N_1+N_2+N_3-1. \label{denstar}
\end{equation}

In this prolate model we take a linear fall to zero in density at the
boundary of the spheroid; then we add an outer partition distance
$R_{s_{N_1+N_2+N_3+1}}$ where $\rho_{N_1+N_2+N_3}^\star[R_{s_{N_1+N_2+
N_3+1}}] = 0$, and the slope of the linear density
$\rho_{N_1+N_2+N_3}^\star[R_S],\ R_{s_{N_1+N_2+N_3}} \leq R_S \leq
R_{s_{N_1+N_2+N_3+1}}$, being the same of $\rho^\star$ in the previous
interval in $R_S$.

For each linear function $\rho_n^\star[R_S]$, the corresponding
density function $\rho_n^\star(a)$ giving the density at the
$unscaled$ variable $a = R_Sa_{x_p} \equiv R_Sa_x$, is
$\rho_n^\star(a) = \rho_n^\star[R_S] = p_{0_n} + p_{1_n} R_S =
p_{0_n} + p_{1_n}a/a_x$. The values of $a$ at the scaled partition
distances $R_{s_n}$ are $a_n = R_{s_n}a_x$.

The mass of the prolate spheroid (= mass of the bar) under the
$\rho^\star$ representation is

\begin{eqnarray} M_{Bar} &=& 4\pi\ (1-e_p^2)\ 
\sum_{n=1}^{N_1+N_2+N_3} \int_{a_n}^{a_{n+1}}  \rho^{\star}_n(a) a^2
da \cr &=& 4\pi\ (1-e_p^2)\sum_{n=1}^{N_1+N_2+N_3}  \left[ \frac{1}{3}
p_{0_n} \left(a_{n+1}^3-a_n^3\right) +  \frac{1}{4} \frac
{p_{1_n}}{a_x} \left(a_{n+1}^4-a_n^4\right) \right].
\label{mbar} \end{eqnarray}

The coefficients $p_{0_n}$ and $p_{1_n}$ are obtained with equations
(\ref{p0n}) and (\ref{p1n}); the evaluation of the density terms in
those equations giving a factor $\rho_0$ (see equation (\ref{denel}),
applied to the prolate spheroid). Then this central density $\rho_0$
will appear as an external factor in equation (\ref{mbar}), from which
it can be obtained once we give $M_{Bar}$;  thus the coefficients
$p_{0_n}, p_{1_n}$ are explicitly known, and so are the functions
$\rho_n^\star(a)$. The set of functions $\rho_n^\star(a)$ is the
function $\rho(a)$ needed in equations (\ref{dpotx}) - (\ref{potpe}).

The integrals in equations (\ref{dpotx}) - (\ref{potpe}) are evaluated
with the intervals in $a_n$, and the corresponding values of
$\beta_p$. For an external point ${\bf r} = (x,y,z)$ the values of
$\beta_p$ at the major semi-axes $a_n$  are the solutions of

\begin{equation} x^2sin^2\beta_{p_n} + R^2\tan^2\beta_{p_n}  = a_n^2e_p^2,\
n=1,2,...,N_1+N_2+N_3+1 . \end{equation}

If the point ${\bf r}$ is inside the prolate spheroid, there is an
$n_0 \in  \{1,2,...,N_1+N_2+N_3\}$ such that  $a_{n_0}\leq a({\bf r})
\leq a_{n_0+1}$ ($a({\bf r})$ is the major semi-axis of the {\it
similar} spheroidal surface passing through the  point ${\bf r}$). In
this case the maximum $\beta_{p_n}$ intervening in the integrals is
$\beta_{p_{n_0}}$ given by $x^2sin^2\beta_{p_{n_0}} +
R^2\tan^2\beta_{p_{n_0}}  = a_{n_0}^2e_p^2$.

Thus, for an internal point ${\bf r}$ equations (\ref{dpotx}) and
(\ref{dpotr}) are

\[ -\frac{\partial\Phi}{\partial x} = -4\pi\ e_p^{-3}\ (1-e_p^2)
x \left\{ \sum_{n=1}^{n_0-1} \int_{\beta_{p_n}}^{\beta{p_{n+1}}}
(p_{0_n}+ \frac {p_{1_n}}{a_x} a(\beta_p)) \frac{sin^2
\beta_p}{cos\beta_p} d\beta_p \right.\]

\begin{equation} \left. + \int_{\beta_{p_{n_0}}}^{sin^{-1}e_p}
(p_{0_{n_0}}+ \frac {p_{1_{n_0}}}{a_x} a(\beta_p)) \frac{sin^2
\beta_p}{cos\beta_p} d\beta_p \right\},
\label{dpotxi} \end{equation}

\[-\frac{\partial\Phi}{\partial R} = -4\pi e_p^{-3} (1-e_p^2) 
R \left\{ \sum_{n=1}^{n_0-1}  \int_{\beta_{p_n}}^{\beta{p_{n+1}}}
\left(p_{0_n}+ \frac {p_{1_n}}{a_x} a(\beta_p)  \right) \frac{sin^2
\beta_p}{cos^3\beta_p} d\beta_p \right.\]

\begin{equation} \left. + \int_{\beta_{p_{n_0}}}^{sin^{-1}\ e_p} 
\left(p_{0_{n_0}}+ \frac {p_{1_{n_0}}}{a_x} a(\beta_p)\right)
\frac{sin^2  \beta_p}{cos^3\beta_p} d\beta_p  \frac{}{}\right\},
\label{dpotri} \end{equation} 

\noindent and the first terms in both equations are excluded if $n_0$
= 1.  The integrals in equations (\ref{dpotxi}) and (\ref{dpotri}) are
analytically easy to find.

The potential at the internal point ${\bf r}$ (see equation
(\ref{potpi})) is

\[ \Phi = -4 \pi e_p^{-1} (1-e_p^2) \left\{ \sum_{n=1}^{n_0-1}
\int_{a_n}^{a_{n+1}} \left(p_{0_n}+ \frac {p_{1_n}}{a_x} a \right) a
\ln\frac{1+sin\beta_p}{cos\beta_p} da + \right.\]

\[ \left. + \int_{a_{n_0}}^{a({\bf r})}\left(p_{0_{n_0}}+ \frac {p_{1_{n_0}}}{a_x} a
\right) a \ln\frac{1+sin\beta_p}{cos\beta_p} da + \right.\]

\begin{equation} \left. + \frac{1}{2}\ln \frac{1+e_p}{1-e_p} \left[
\int_{a({\bf r})}^{a_{n_0+1}}\left(p_{0_{n_0}}+ \frac
{p_{1_{n_0}}}{a_x} a \right) a~da + \sum_{n=n_0+1}^{N_1+N_2+N_3}
\int_{a_n}^{a_n+1}\left(p_{0_{n_0}}+ \frac {p_{1_{n_0}}}{a_x} a
\right) a~da \right] \right\}, \label{poti} \end{equation}

\noindent and we exclude the first or last sums  if $n_0$ = 1 or $n_0
= N_1+N_2+N_3$, respectively.

With $I_n(a) = \int_{0}^{a} \rho^{\star}_n(u) udu = \frac
{1}{2}p_{0_n}a^2 + \frac {1}{3} \frac {p_{1_n}}{a_x}a^3$, and
integrating by parts, equation (\ref{poti}) can be written as

\[ \Phi = -4 \pi e_p^{-1} (1-e_p^2) \left\{ -\sum_{n=1}^{n_0-1}
\int_{\beta_{p_n}}^{\beta_{p_{n+1}}} I_n(a(\beta_p)) \frac
{d\beta_p}{cos\beta_p}- \int_{\beta_{p_{n_0}}}^{sin^{-1} \ e_p}
I_{n_0}(a(\beta_p)) \frac {d\beta_p}{cos\beta_p} + \right.\]

\[ \left. + \sum_{n=1}^{n_0-1} \left[ I_n(a_{n+1})\ln\frac{1+sin\beta_{p_{n+1}}}
{cos\beta_{p_{n+1}}} -
I_n(a_n)\ln\frac{1+sin\beta_{p_n}}{cos\beta_{p_n}} \right] + \right.\]

\begin{equation} \left. + \frac{1}{2}\ln \frac{1+e_p}{1-e_p} \left[ I_{n_0}(a_{n_0+1}) +
\sum_{n=n_0+1}^{N_1+N_2+N_3} \left[ I_n(a_{n+1}) - I_n(a_n) \right]
\right] -
I_{n_0}(a_{n_0})\ln\frac{1+sin\beta_{p_{n_0}}}{cos\beta_{p_{n_0}}}
\right\}, \label{potii} \end{equation}

\noindent with the explicit form of $I_n(a)$, the integrals in this
equation are also easy to solve analytically.

For an external point ${\bf r}$ the accelerations and potential are

\begin{equation} -\frac{\partial\Phi}{\partial x} =
-4\pi\ e_p^{-3}\ (1-e_p^2) x \sum_{n=1}^{N_1+N_2+N_3}
\int_{\beta_{p_n}}^{\beta_{p_{n+1}}} (p_{0_n}+ \frac {p_{1_n}}{a_x}
a(\beta_p)) \frac{sin^2 \beta_p}{cos\beta_p} d\beta_p,
\label{dpotxe} \end{equation}

\begin{equation} -\frac{\partial\Phi}{\partial R} = -4\pi e_p^{-3} 
(1-e_p^2) R \sum_{n=1}^{N_1+N_2+N_3}
\int_{\beta_{p_n}}^{\beta_{p_{n+1}}} (p_{0_n}+ \frac {p_{1_n}}{a_x}
a(\beta_p))  \frac{sin^2 \beta_p}{cos^3\beta_p} d\beta_p,
\label{dpotre} \end{equation} 

\[ \Phi = -4 \pi e_p^{-1} (1-e_p^2) \left\{ -\sum_{n=1}^{N_1+N_2+N_3}
\int_{\beta_{p_n}}^{\beta_{p_{n+1}}} I_n(a(\beta_p)) \frac
{d\beta_p}{cos\beta_p}+ \right.\]

\begin{equation} \left. + \sum_{n=1}^{N_1+N_2+N_3} \left[ I_n(a_{n+1})
\ln\frac{1+sin\beta_{p_{n+1}}} {cos\beta_{p_{n+1}}} -
I_n(a_n)\ln\frac{1+sin\beta_{p_n}}{cos\beta_{p_n}} \right]
\right\}. \label{pote} \end{equation}

\section{The Model of Superposition of Ellipsoids}\label{msuperp}

One of the most conspicuous characteristics of the Galactic bar  (also
observed in other galaxies) is the ``boxy'' form of its isophotes
(Freudenreich 1998; Zhao \& Mao 1996; Ibata \& Gilmore 1995).  In the
ellipsoidal and prolate models of the Galactic bar the iso-density
contours  are elliptical. In order to approximate the boxy iso-density
contours  we have considered a superposition of inhomogeneous
ellipsoids.

\begin{figure}
\begin{center}
\epsscale{0.5}
\plotone{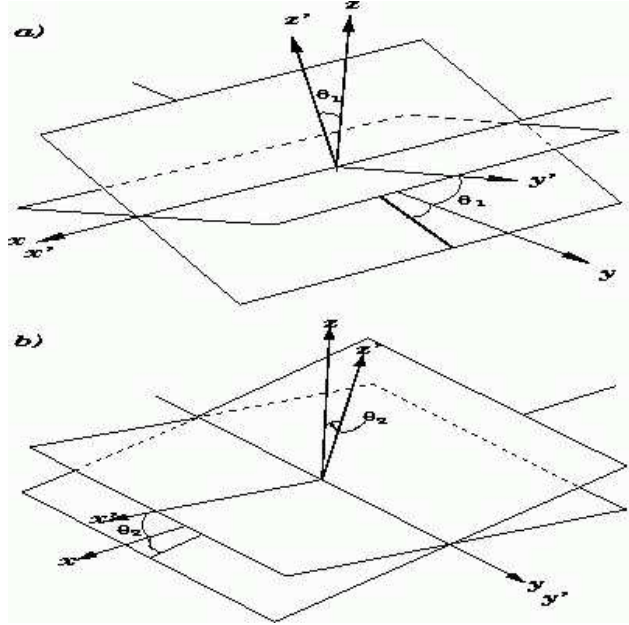}                      
\caption {The planes that contain the major and middle axes of the
four ellipsoids in our third model. In ($a$) two identical ellipsoids
have their major axes, $x'$, along the $x$-axis and their  middle,
$y'$, and minor, $z'$, axes rotated an angle $\theta_1$ around the
$x$-axis, to both sides of the $z$-axis. In ($b$) we show the
orientation of the other two ellipsoids, with their middle axes, $y'$,
along the $y$-axis and their major, $x'$, and minor, $z'$, axes
rotated an angle $\theta_2$ around the $y$-axis, to both sides of the
$z$-axis. Our model is the superposition of the ellipsoids in ($a$)
and ($b$).}
\label{fig.superp}
\end{center}
\end{figure}

The model consists of four ellipsoids with a density of the functional
form given  in equation (\ref{denel}). The $x,y,z$-axes define the
major, middle, and minor axes of the Galactic bar. Two identical
ellipsoids have their  major axes along the $x$-axis, and their middle
and minor axes  are rotated around the $x$-axis an angle $\theta_1$ to
both sides of  the $z$-axis (see Figure \ref{fig.superp} $a$).  The
other two ellipsoids, also identical, have their middle axes along the
$y$-axis, and their major and minor axes  are rotated around the
$y$-axis an angle $\theta_2$ to both sides of the $z$-axis (see
Figure \ref{fig.superp} $b$). The reason to try this arrangement of
ellipsoids is simply that it favors the boxy edge-on appearance of the
bar.

Our task is to find the dimensions, relative masses, and orientation
angles $\theta_1, \theta_2$ of the four ellipsoids, such that their
superposition gives a good approximation to MSF.

The ellipsoidal model in Section \ref{modelip} has the effective
semi-axes $a_{Bar}$, $b_{Bar} = a_{Bar}a_y$, $c_{Bar} = a_{Bar}a_z$,
and corresponding scale lengths $a_x, a_y, a_z$.  In its outer region
the Gaussian factor has the scale length $h_{end}$. For the two
identical ellipsoids rotated the angle $\theta_1$ (see
Fig. \ref{fig.superp} $a$), we take their dimensions and scale lengths
as $a_{Bar_1} = k_1a_{Bar},\ b_{Bar_1} = k_2b_{Bar},\ c_{Bar_1} =
k_3c_{Bar},\ a_{x1} = k_1a_x,\ a_{y1} = k_2a_y,\ a_{z1} = k_3a_z,\
h_{end1} = k_1h_{end}$, and each ellipsoid with a mass $M_1$, which is
a fraction $k_4$ of the total mass of the bar (= sum of masses of the
four ellipsoids in the model): $M_1 = k_4M_{Bar}$. Likewise, for the
other two identical ellipsoids rotated the angle $\theta_2$ (see
Fig. \ref{fig.superp} $b$), the dimensions, scale lengths, and masses
are taken as $a_{Bar_2} = k_5a_{Bar},\ b_{Bar_2} = k_6b_{Bar},\
c_{Bar_2} = k_7c_{Bar},\ a_{x2} = k_5a_x,\ a_{y2} = k_6a_y,\ a_{z2} =
k_7a_z,\ h_{end2} = k_5h_{end},\ M_2 = k_8M_{Bar}$. $k_1,k_2,...,k_8$
are positive constants. The constraints are $2k_4 + 2k_8$ = 1 and
$a_{Bar_1} > b_{Bar_1} > c_{Bar_1},\ a_{Bar_2} > b_{Bar_2} >
c_{Bar_2}$, i.e., $k_1/k_2 > a_y/a_x,\ k_2/k_3 > a_z/a_y,\ k_5/k_6 >
a_y/a_x,\ k_6/k_7 > a_z/a_y$.

The density of each ellipsoid in the superposition has the form given
in equation (\ref{denel}), using the corresponding dimensions, scale
lengths, and total mass of the ellipsoid. This density is computed in
a Cartesian system whose axes are the principal axes of the
ellipsoid. The required central density $\rho_0$ is obtained from an
equation analogous to equation (\ref{melip}), using a fine partition
$(N_1,N_2,N_3)$ (we take the same partition in the four ellipsoids of
the model) and with the corresponding mass of the ellipsoid.

Thus, formally, with the dimensions, mass, and orientation of each
ellipsoid, we can write the expression for the density due to the
superposition, $\rho_{sup}({\bf r})$, at any point ${\bf r}=(x,y,z)$
in the Cartesian system whose axes $x,y,z$ are the principal axes of
the bar (i.e., the $x$- and $y$- axes lie on the Galactic plane, and
the $z$-axis is the rotation axis of the Galaxy).

Once the numbers $k_1,k_2,...,k_8$ and $\theta_1, \theta_2$ are
chosen, the surfaces $\rho_{sup}({\bf r})/\rho_{sup}(0) = c =
constant$  can be obtained, and thus compared with the corresponding
surfaces $\rho_S({\bf r})/\rho_S(0) = c =\ constant$  arising from
MSF, whose density is of the type in equation (\ref{denel}), but with
$R_S$ given by equation (\ref{Rsfre}).

The procedure to obtain a fit to MSF by means of the proposed
superposition of ellipsoids is as follows: due to the symmetry of the
surfaces $\rho_{sup}({\bf r})/\rho_{sup}(0) = c$ and $\rho_S({\bf
r})/\rho_S(0) = c$, we only consider the octant $x\geq 0$, $y\geq 0$,
$z\geq 0$. In this octant we take a fine mesh of radial directions
given by spherical angles $(\varphi,\theta)$ (the $z$-axis being the
polar axis). In the mesh points $(\varphi,\theta)_i$ we find the
distances to the origin, $r_{sup},\ r_S$, of the corresponding points
on the surfaces $\rho_{sup}({\bf r})/\rho_{sup}(0) = c$ and
$\rho_S({\bf r})/\rho_S(0) = c$.  Thus, for $n$ mesh points and $m$
values of the constant $c$ (i.e., comparing $m$ pairs of surfaces), we
minimize the quantity

\begin{equation} D=\left\{\frac{\sum_{j=1}^m\frac{c_j}{n}
\sum_{i=1}^n(r_S-r_{sup})^2_{ij}}{\sum_{j=1}^mc_j}\right\}^{1/2},
\label{Derror}\end{equation}

\noindent which gives an $rms$ separation in all the pairs of surfaces.

$D$ is a function of the ten variables
$k_l,k_2,...,k_8,\theta_1,\theta_2$.  We have employed the algorithm
$amoeba$ of Press \etal (1992) to minimize the function $D$ in the
10-dimensional space $(k_1,k_2,...,k_8,\theta_1,\theta_2)$, under the
imposed constraints on the variables $k_1,k_2,...,k_8$.  The point
$(k_1,k_2,...,k_8,\theta_1,\theta_2)$ where  the function $D$ reaches
its minimum value gives our best fit to MSF.  We obtain the following
values of the variables at this point

\begin{eqnarray} 
k_1&=&1.1982\cr k_2&=&1.4086\cr k_3&=&0.8565\cr k_4&=&0.2600\cr
\theta_1&=&0.6758\ {\rm rad}\cr k_5&=&1.1803\cr k_6&=&1.1941\cr
k_7&=&1.0123\cr k_8&=&0.2400\cr \theta_2&=&0.2420\ {\rm rad}
\nonumber\end{eqnarray}

With all these properties of the ellipsoids in the superposition, the
potential and force field of each ellipsoid in its corresponding
Cartesian axes are obtained with the procedure given in Section
\ref{modelip} for the ellipsoidal model. Thus, the potential and force
field are obtained at any point ${\bf r}=(x,y,z)$ in the Cartesian
system of the principal axes of the bar.

\begin{figure}
\begin{center}
\epsscale{1.0} \plotone{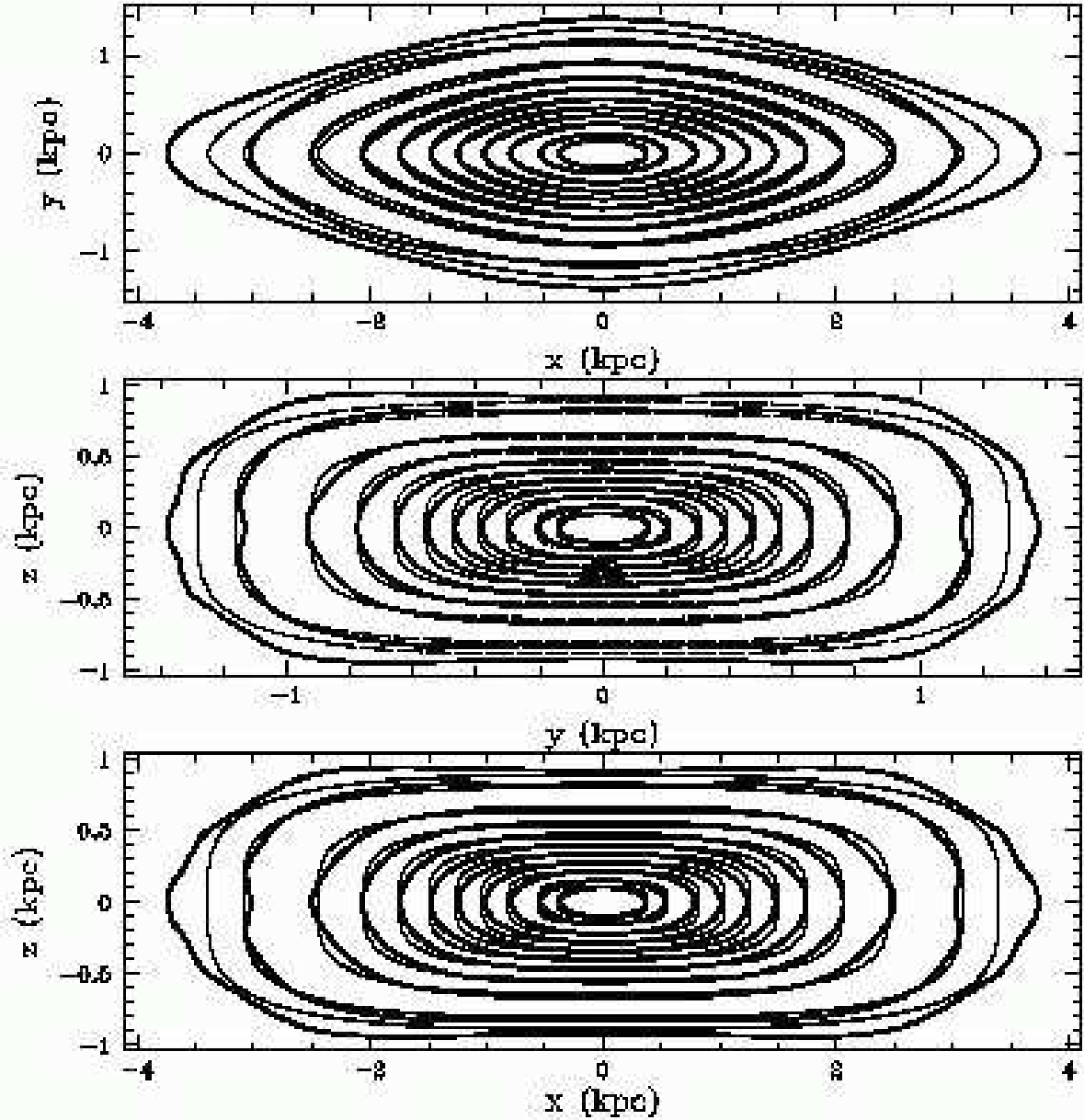}
\caption {Iso-density contours on the $x-y$  plane -the Galactic
plane- (top panel), $y-z$ plane (middle panel),  and $x-z$ plane
(bottom panel) of our third model for the Galactic  bar (dark lines),
and of Model S of Freudenreich (1998) (MSF)  (light lines). The
comparison is made for values of the constant $c$ (see text) = 0.95,
0.9, 0.8, 0.7, 0.6, 0.5, 0.4, 0.3, 0.2, 0.1, and 0.05 . It is worth
noticing that the scale in the three figures  is not the same.}
\label{fig.denxy}
\end{center}
\end{figure}

\begin{figure}
\begin{center}
\epsscale{1.0} \plotone{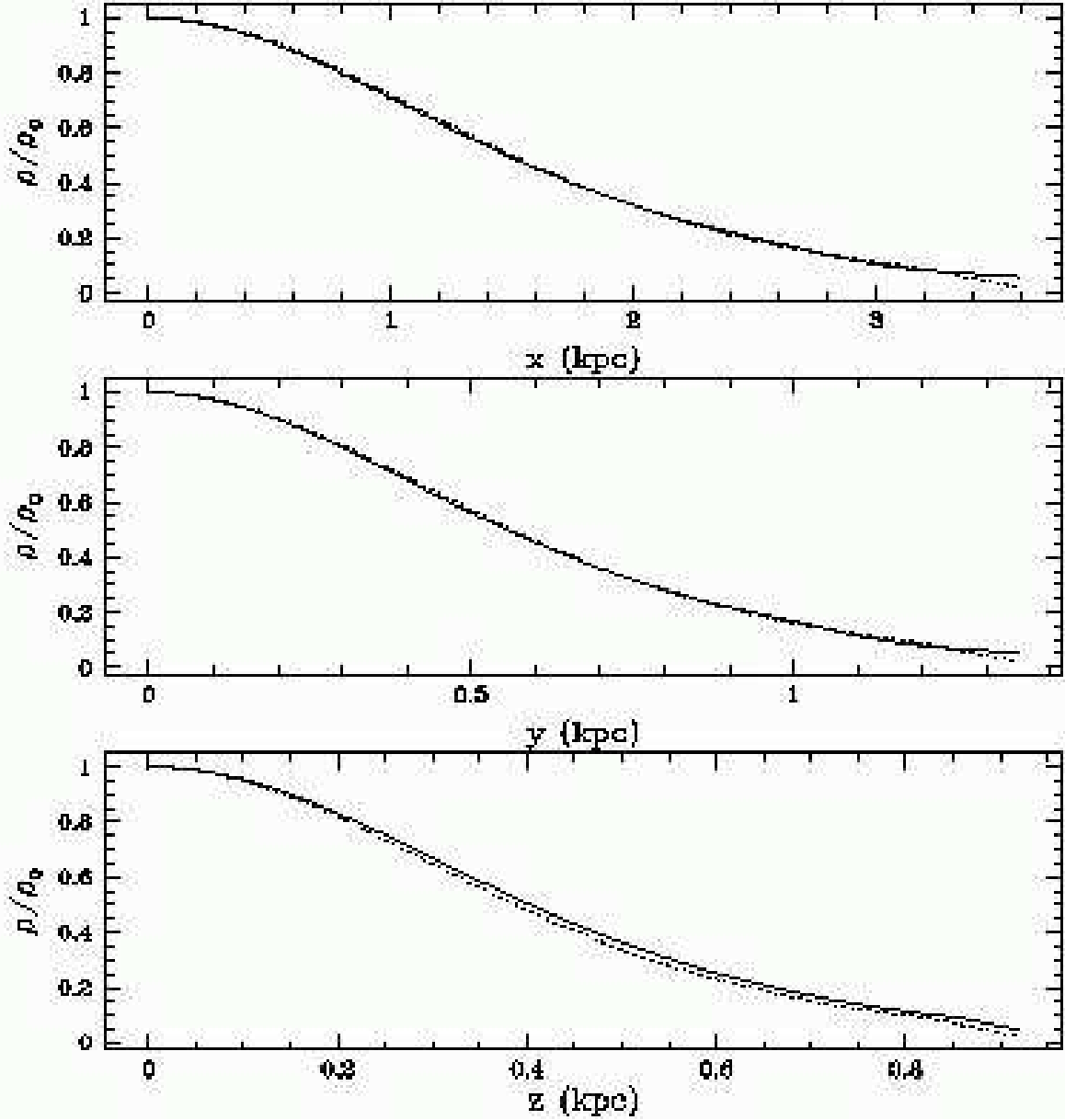}
\caption {Density along the $x$-axis (top panel), $y$-axis  (middle
panel), and  $z$-axis (bottom panel) -which represent  the major,
middle, and minor axes, respectively, of the bar- in our third model for
the Galactic bar (continuous line), and for Model S of Freudenreich
(1998) (MSF) (dotted lines).}
\label{fig.denx}
\end{center}
\end{figure}

In figure \ref{fig.denxy} we show some iso-density contours on the
three Cartesian planes, obtained with our superposition model (dark
lines), and with MSF (light lines). Figure \ref{fig.denx} shows the
density of our model (continuous lines), and of MSF (dotted lines),
along the three principal axes of the bar.

\begin{figure}
\begin{center}
\epsscale{1.0} \plotone{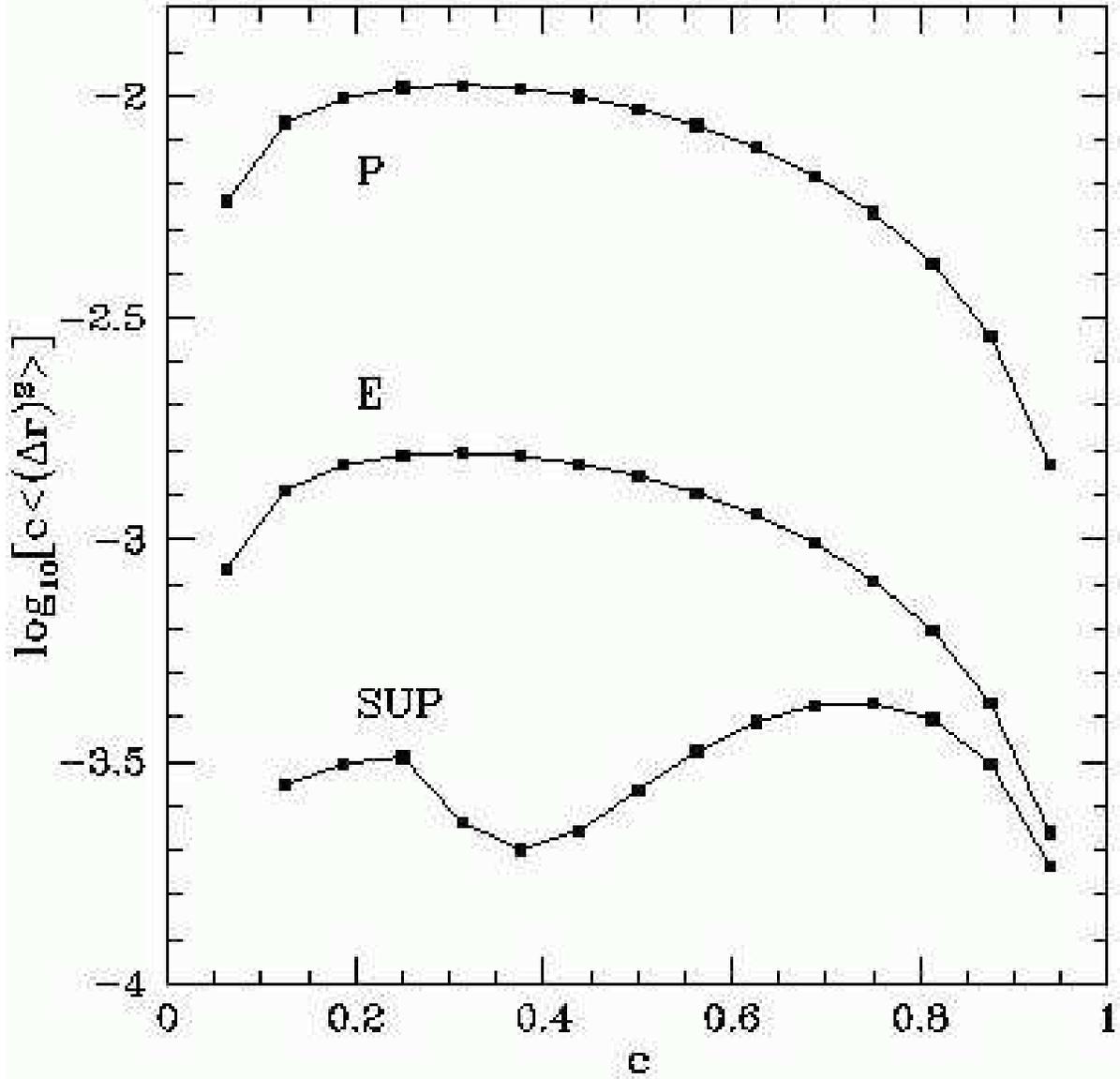}
\caption {The weighted mean-squared separation between iso-density
surfaces, $\rho/\rho_0$ = c, in Model S of Freudenreich (1998, MSF)
and  corresponding surfaces in our three models (E: ellipsoidal, P:
prolate, SUP: superposition of ellipsoids).}
\label{fig.promsup}
\end{center}
\end{figure}

To see how MSF is approximated by the three proposed models of the
Galactic bar, we can compare the weighted mean-squared separation
$c_j<(\Delta r)^2> \equiv
\frac{c_j}{n}\sum_{i=1}^n(r_S-r_{mod})^2_{ij}$ between iso-density
surfaces $\rho_S({\bf r})/\rho_S(0) = c_j$, \ $\rho_{mod}({\bf
r})/\rho_{mod}(0) = c_j$, where $mod$ indicates any of the three
models. Figure \ref{fig.promsup} shows this comparison. Clearly, our
third model gives a better approximation to MSF.

\section{Analysis of Force Fields}\label{analysis}

In this section we give a brief analysis of the dependence of the
force field obtained with the ellipsoidal, prolate, and superposition
models (Sections \ref{modelip}, \ref{mprol},  and \ref{msuperp} of
this Appendix) on the values of the partition numbers $N_1, N_2, N_3$.
In the ideal limit these must be very large numbers, but it is
important to find in each model a representative set $(N_1,N_2,N_3)$
which shall give a gravitational field very similar to that obtained
with large numbers, and not being excessively time-consuming in
numerical orbital integrations.

Some tests showed that in the ellipsoidal and superposition models a
partition with $N_1+N_2+N_3 \sim 100$ satisfy the time requirement,
and gives a small  rate of change of the force field under variations
of $N_1,N_2,N_3$. For the  prolate model, an appropriate partition has
$N_1+N_2+N_3 \sim 30$; this reduction of  components is expected since
the linear functions employed in this model approximate the density
function better, and a small number suffices.

To analyze in detail if the proposed values of $N_1+N_2+N_3$ are
indeed appropriate, we have taken the definite numbers $N_1 = 20,\ N_2
= 65,\ N_3 = 15$ in the ellipsoidal and superposition models, and $N_1
= 15,\ N_2 = 20,\ N_3 = 5$ in the prolate model, with $a_L = 1$ kpc
(see Section \ref{modelip}) in all cases. Also, we take the mass of
the bar as $9.8 \times 10^{9}$ M$\odot$ (see Section \ref{fit_obs}).

\begin{figure}
\begin{center}
\epsscale{1.0} \plotone{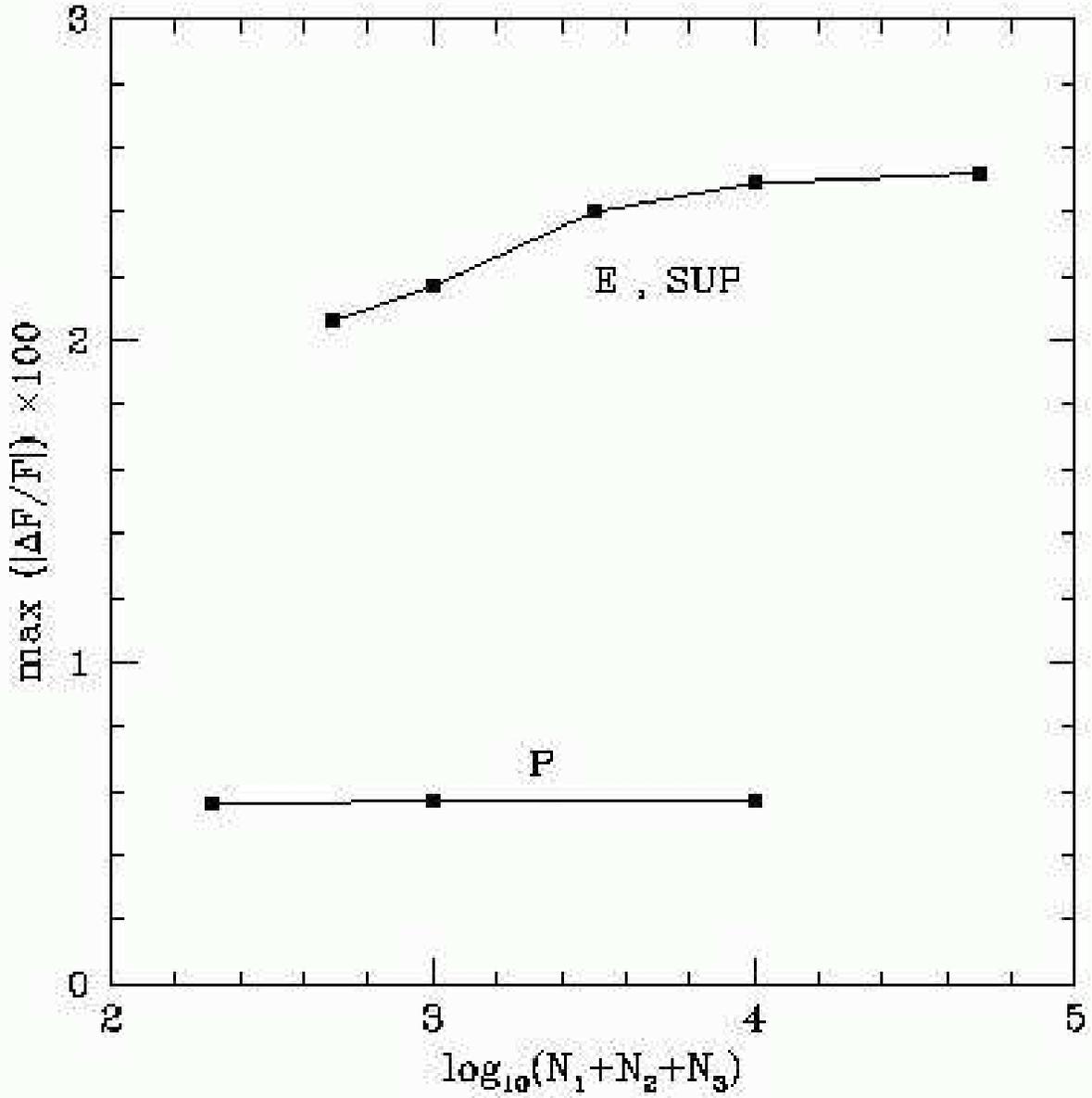}
\caption {Percentage force difference comparing the force obtained
with the partition numbers $(N_1,N_2,N_3)_0 = (20,65,15)$ in the
ellipsoidal (E) and superposition (SUP) models, $(N_1,N_2,N_3)_0 =
(15,20,5)$ in the prolate (P) model, and the force obtained with
increasing partition numbers.}
\label{fig.porcent}
\end{center}
\end{figure}

\begin{figure}
\begin{center}
\epsscale{1.0} \plotone{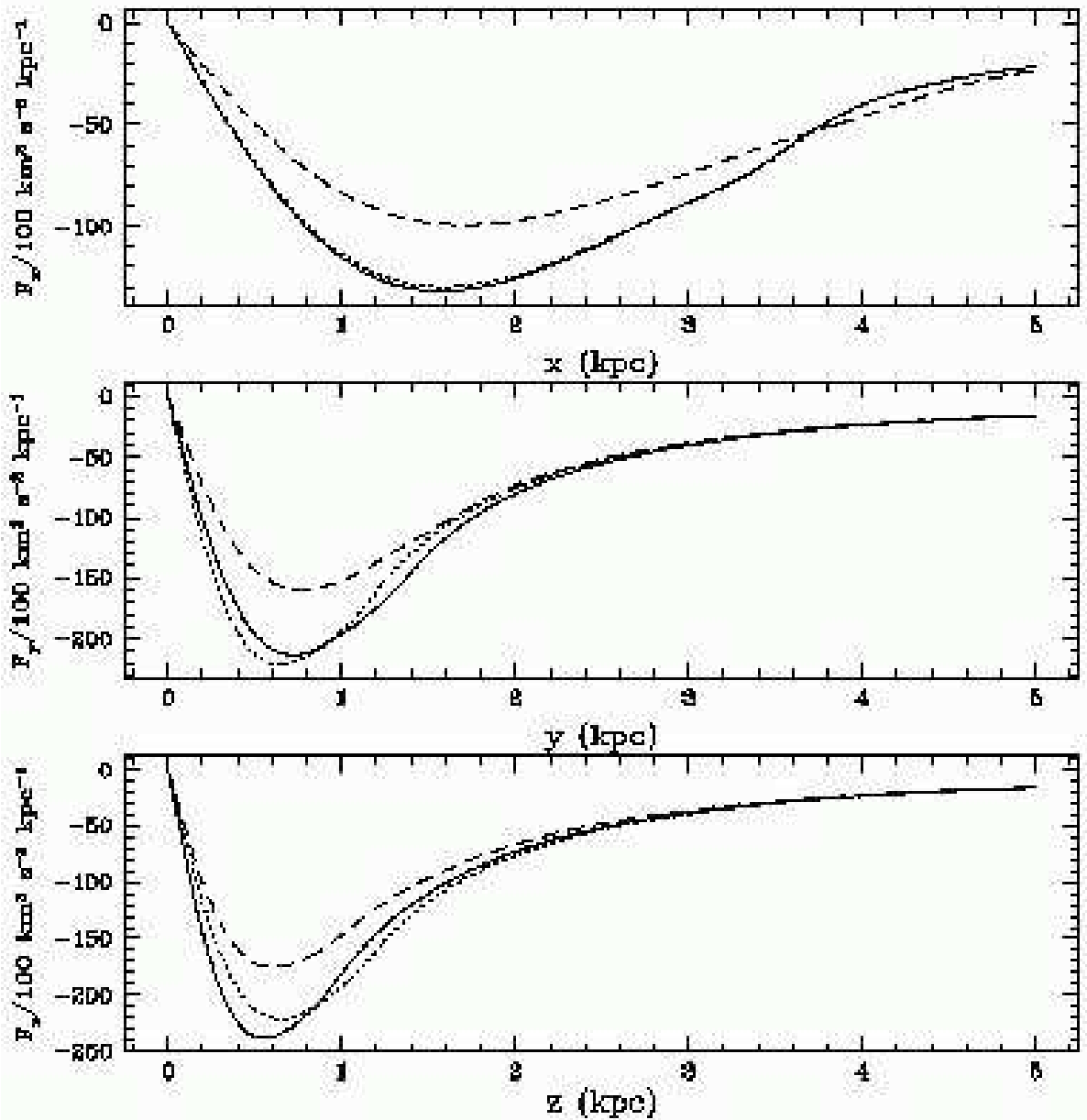}
\caption {Acceleration along the $x$-axis (top panel), $y$-axis 
(middle panel), and $z$-axis (bottom panel), with the partition 
$(N_1,N_2,N_3)_0 = (20,65,15)$ in the ellipsoidal (continuous 
line) and superposition (discontinuous line) models, and with 
$(N_1,N_2,N_3)_0 = (15,20,5)$ in the prolate model (dotted line).}
\label{fig.fx}
\end{center}
\end{figure}

Keeping $a_L$ fixed, we increase the partition numbers by a certain
factor and compute the resulting force (acceleration) $F$ along the
principal axes of the bar.  This force is compared at corresponding
distances on a given axis with the force $F_0$ obtained using the
numbers given above, and we find the maximum relative difference
$|\Delta F/F| = |(F - F_0)/F|$ from the three axes. Figure
\ref{fig.porcent} shows the results of this analysis. The conclusion
from this figure is that the partitions $(N_1,N_2,N_3)_0 = (20,65,15)$
in the ellipsoidal and superposition models, and $(N_1,N_2,N_3)_0 =
(15,20,5)$ in the prolate model, give a  force field, ${\bf F_0}$,
which differs $\sim 2.5 \%$ and $0.6 \%$,  respectively, from that
obtained in the ideal limit of very large  partition numbers. Finally,
Figure \ref{fig.fx} shows the acceleration along the three principal
axes obtained with the partitions $(N_1,N_2,N_3)_0$. The superposition
model differs significantly from the ellipsoidal and prolate
models.

\end{appendix}

%%%%%%%%%%%%%%%%%%%%%%%%%%%%%%%%%%%%%%%%%%%%%%%%%%%%%%%%%%%%%%%%%%%%%%%

\clearpage

\begin{tabular}{cr@{.}lr@{.}lr@{.}lr@{.}lr@{.}lr@{.}l}
\multicolumn{13}{c}{TABLE 1} \\
\multicolumn{13}{c}{Orbital Properties in the Axisymmetric Galactic Model} \\
\multicolumn{13}{c}{of Allen \& Santill\'an (1991)} \\ \hline\hline
NGC & \multicolumn{2}{c}{$<r_{min}>$} & \multicolumn{2}{c}{$<r_{max}>$} &
\multicolumn{2}{c}{$<|z|_{max}>$} & \multicolumn{2}{c}{$<e>$} &
\multicolumn{2}{c}{$E$} & \multicolumn{2}{c}{$h$} \\ \hline
5139 & 1 & 11 & 6 & 97 & 0 & 69 & 0 & 72 & -1492 & 5 & -42 & 4 \\
6093 & 0 & 58 & 3 & 67 & 1 & 52 & 0 & 73 & -1713 & 5 & -6 & 0 \\
6144 & 2 & 22 & 2 & 82 & 2 & 48 & 0 & 12 & -1663 & 9 & -20 & 3 \\
6171 & 2 & 82 & 3 & 79 & 2 & 23 & 0 & 15 & -1578 & 2 & 46 & 9 \\
6218 & 3 & 02 & 5 & 74 & 2 & 44 & 0 & 31 & -1458 & 9 & 64 & 3 \\
6712 & 1 & 14 & 6 & 75 & 1 & 93 & 0 & 71 & -1478 & 7 & 22 & 3 \\ \hline
\multicolumn{13}{l}{Distances in kpc; $E$ in units of 100 km$^2$ s$^{-2}$; $h$ in units
of} \\
\multicolumn{13}{l}{10 km s$^{-1}$ kpc.} \\
\end{tabular}

\clearpage

\begin{tabular}{cr@{.}lr@{.}lr@{.}lr@{.}lr@{.}lr@{.}lr@{.}lr@{.}l}
\multicolumn{17}{c}{TABLE 2} \\
\multicolumn{17}{c}{Orbital Properties with the three Models of the Galactic Bar}
\\ \hline\hline
NGC & \multicolumn{2}{c}{$<r_{min}>$} & \multicolumn{2}{c}{$<r_{max}>$} &
\multicolumn{2}{c}{$<|z|_{max}>$} & \multicolumn{2}{c}{$<e>$} &
\multicolumn{2}{c}{$E_{min}$} & \multicolumn{2}{c}{$E_{max}$} &
\multicolumn{2}{c}{$h_{min}$} & \multicolumn{2}{c}{$h_{max}$} \\ \hline
5139 & 1 & 62 & 6 & 37 & 1 & 71 & 0 & 58 & -1554 & 6 & -1375 & 3 & -52 & 7 &
-22 & 8 \\
     & 1 & 01 & 8 & 14 & 1 & 41 & 0 & 77 & -1561 & 9 & -1314 & 6 & -53 & 9 &
-12 & 7 \\
     & 1 & 15 & 8 & 39 & 2 & 42 & 0 & 76 & -1538 & 4 & -1304 & 3 & -50 & 0 &
-11 & 0 \\ \hline
6093 & 0 & 69 & 4 & 00 & 1 & 50 & 0 & 69 & -1851 & 9 & -1523 & 7 & -30 & 4 &
24 & 3 \\ 
     & 0 & 81 & 4 & 00 & 1 & 61 & 0 & 65 & -1769 & 8 & -1550 & 3 & -16 & 6 &
20 & 0 \\
     & 1 & 55 & 2 & 90 & 2 & 52 & 0 & 31 & -1753 & 7 & -1664 & 8 & -14 & 4 &
0 & 4 \\ \hline
6144 & 2 & 30 & 3 & 23 & 2 & 90 & 0 & 17 & -1675 & 0 & -1567 & 5 & -23 & 8 &
-5 & 9 \\
     & 1 & 83 & 3 & 45 & 2 & 57 & 0 & 31 & -1715 & 0 & -1551 & 4 & -30 & 4 &
-3 & 1 \\
     & 1 & 98 & 3 & 46 & 2 & 78 & 0 & 28 & -1684 & 5 & -1563 & 1 & -25 & 9 &
-5 & 7 \\ \hline
6171 & 1 & 57 & 3 & 65 & 1 & 49 & 0 & 42 & -1843 & 1 & -1482 & 9 & 3 & 3 &
63 & 4 \\
     & 2 & 00 & 3 & 58 & 1 & 83 & 0 & 30 & -1838 & 3 & -1451 & 4 & 4 & 2 &
68 & 7 \\
     & 2 & 03 & 3 & 24 & 1 & 95 & 0 & 22 & -1793 & 6 & -1455 & 2 & 11 & 7 &
68 & 1 \\ \hline
6218 & 2 & 16 & 4 & 51 & 2 & 01 & 0 & 36 & -1783 & 9 & -1415 & 2 & 10 & 7 &
72 & 2 \\
     & 2 & 94 & 5 & 93 & 2 & 35 & 0 & 34 & -1521 & 4 & -1378 & 8 & 54 & 5 &
78 & 3 \\
     & 2 & 23 & 5 & 11 & 1 & 89 & 0 & 40 & -1750 & 1 & -1406 & 4 & 16 & 6 &
73 & 9 \\ \hline
6712 & 1 & 03 & 6 & 14 & 2 & 39 & 0 & 70 & -1595 & 8 & -1353 & 9 & 4 & 1 &
44 & 4 \\
     & 1 & 00 & 6 & 72 & 1 & 72 & 0 & 75 & -1603 & 4 & -1401 & 5 & 2 & 8 &
36 & 5 \\
     & 0 & 98 & 6 & 60 & 1 & 77 & 0 & 74 & -1601 & 5 & -1418 & 7 & 3 & 5 &
34 & 0 \\ \hline
\multicolumn{17}{l}{-Distances in kpc; $E$ in units of 100 km$^2$ s$^{-2}$;
$h$ in units of 10 km s$^{-1}$ kpc.} \\
\multicolumn{17}{l}{-Lines in each entry: first: ellipsoidal model, second:
prolate model,} \\
\multicolumn{17}{l}{\ third: superposition model.} \\
\end{tabular}

\clearpage

\clearpage

\clearpage

\clearpage

\clearpage

\clearpage

\clearpage

\clearpage

\clearpage

\clearpage

\clearpage

\clearpage

\clearpage

\clearpage

\clearpage

\clearpage

%\newpage

%\label{lastpage}

\end{document}